\documentclass[aps,prl,twocolumn,superscriptaddress,showpacs,floatfix,longbibliography]{revtex4-2}  
\usepackage{amsmath,amssymb,wasysym,graphicx}
\usepackage{times}
\usepackage[varg]{txfonts}
\usepackage{textcomp}
\usepackage{subfigure}
\usepackage{tabu}
\usepackage{color}
\usepackage[colorlinks=true,citecolor=blue,urlcolor=blue,linkcolor=blue,hyperindex]{hyperref}
\usepackage{braket}
\usepackage{overpic}
\usepackage{clrscode3e} 
\newcommand{\bl}[1]{{\color{blue}#1}} 
\newcommand{\rd}[1]{{\color{red}#1}} 
\usepackage[normalem]{ulem}
\usepackage{verbatim}
\allowdisplaybreaks

\begin{document}
	\title{Amplitude Mode in Quantum Magnets via Dimensional Crossover}
	
	\author{Chengkang Zhou}
	\affiliation{Department of Physics and HKU-UCAS Joint Institute of Theoretical and Computational Physics, The University of Hong Kong, Pokfulam Road, Hong Kong SAR, China}
	
	\author{Zheng Yan}
	\affiliation{Department of Physics and HKU-UCAS Joint Institute of Theoretical and Computational Physics, The University of Hong Kong, Pokfulam Road, Hong Kong SAR, China}
	\affiliation{State Key Laboratory of Surface Physics and Department of Physics, Fudan University, Shanghai 200438, China}
	
	\author{Han-Qing Wu}
	\affiliation{School of Physics, Sun Yat-Sen University, Guangzhou, 510275, China}
	
	\author{Kai Sun}
	\email{sunkai@umich.edu}
	\affiliation{Department of Physics, University of Michigan, Ann Arbor, Michigan 48109, USA}
	
	\author{Oleg A. Starykh}
	\email{starykh@physics.utah.edu}
	\affiliation{Department of Physics and Astronomy, University of Utah, Salt Lake City, Utah 84112, USA}

	\author{Zi Yang Meng}
	\email{zymeng@hku.hk}
	\affiliation{Department of Physics and HKU-UCAS Joint Institute of Theoretical and Computational Physics, The University of Hong Kong, Pokfulam Road, Hong Kong SAR, China}
	
	\date{\today}
	
	\begin{abstract}
		We investigate the amplitude (Higgs) mode associated with longitudinal fluctuations of the order parameter at the continuous spontaneous symmetry breaking phase transition. 
		In quantum magnets, due to the fast decay of the amplitude mode into low-energy Goldstone excitations, direct observation of this mode represents a challenging task. 
		By focusing on a quasi-one-dimensional geometry, we circumvent the difficulty and investigate the amplitude mode in a system of weakly coupled spin chains with the help of 
		quantum Monte Carlo simulations, stochastic analytic continuation, and a chain-mean field approach combined with a mapping to the field-theoretic sine-Gordon model.
		The amplitude mode is observed to emerge in the longitudinal spin susceptibility in the presence of a weak symmetry-breaking staggered field. A conventional measure of the amplitude mode in higher dimensions,  the singlet bond mode, is found to appear at a lower than the amplitude mode frequency. We identify these two excitations with the second (first) breather of the sine-Gordon theory, correspondingly. In contrast to higher-dimensional systems, the amplitude and bond order fluctuations are found to carry significant spectral weight in the quasi-1D limit.
	\end{abstract}
	\maketitle
	
	{\noindent \it Introduction.-----} The phenomenon of spontaneous symmetry breaking (SSB) represents one of the key notions in modern physics.
	For a continuous global symmetry, SSB is expected to generate two types of collective excitations -- Goldstone modes, describing transverse or phase fluctuations of the order parameter, and Higgs modes, which describe its longitudinal or amplitude fluctuations.
	In contrast to the gapless Goldstone excitation, which is commonly observed in a variety of condensed matter systems (e.g., magnons in magnetically ordered materials), the observation of the amplitude (longitudinal) mode is more challenging. It is complicated by its intrinsically finite lifetime -- an amplitude-mode excitation is allowed to decay into a pair of Goldstone excitations which leads to a strong damping of this excitation. By now several successful experimental sightings of the amplitude mode have been reported in the dimerized ~\cite{Merchant2014} and quasi-one-dimensional (1D) quantum magnets $\mathrm{KCuF_3}$~\cite{Novel2000Lake,Longitudinal2005Lake}, 
	$\mathrm{BaCu_2Si_2O_7}$~\cite{Spin2001Zheludev}, Ising-like spin chains SrCo$_2$V$_2$O and Yb$_2$Pt$_2$Pb~\cite{Bera2017,Gannon2019} as well as in superconducting settings ~\cite{Sherman2015,Shimano2020}. 
	
	The amplitude mode is a well-defined excitation when its lifetime is long, which requires suppression of the decays into Goldstone modes, the spin waves. 
	Theoretically, such suppression requires weakening of the long range magnetic order, the magnitude of which determines the spectral weight of the spin waves.
	Two ways to achieve this have been proposed, through (a) quantum critical points (QCPs)~\cite{Sachdev1999,Podolsky2011,Gazit2013,Sushkov2017} and
	(b) dimensional crossover towards one dimension (1D)~\cite{Raman1992Canali,Longitudinal1992Ian,schulz1996,Quasi-one-dimensional1997Essler}. 
	The first strategy was recently verified via quantum Monte Carlo model simulations in dimerized antiferromagnet~\cite{Dynamical2015Lohofer,Amplitude2017Qin,Lohoefer2017} and superconductor-insulator transition~\cite{Swanson2014}.
	
	\begin{figure}[htp!]
		\centering
		\includegraphics[width=0.55\columnwidth]{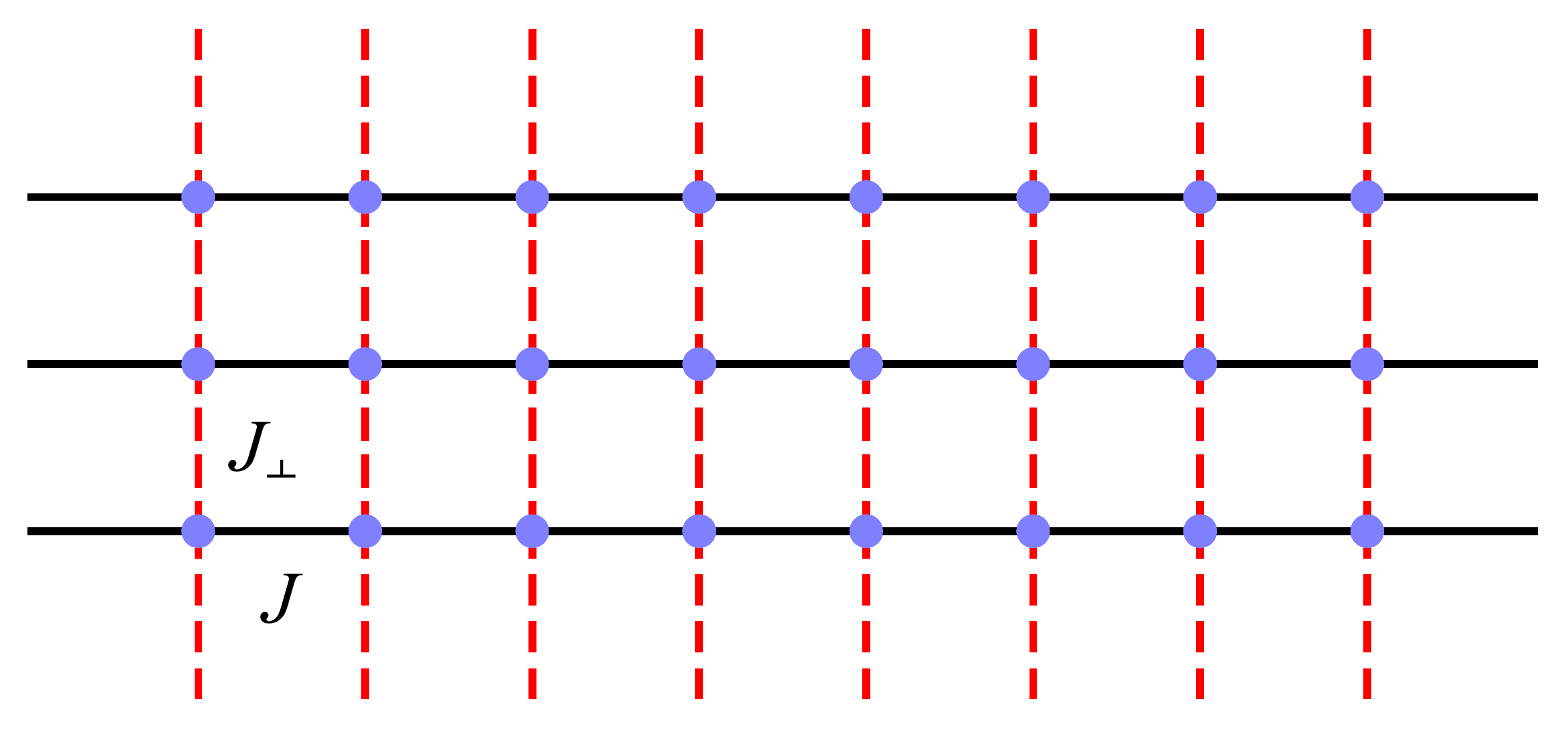}
		\caption{Coupled antiferromagnetic Heisenberg spin chains with nearest-neighbor spin exchange $J$ (black solid line) and $J_{\perp}$ (red dashed line).}
		\label{fig:fig1}
	\end{figure}
	
	In this Letter, we explore the second, quasi-1D approach. It has long been proposed that a stable longitudinal model shall arise in weakly coupled spin chains~\cite{Longitudinal1992Ian,schulz1996,Quasi-one-dimensional1997Essler}. 
	It should be noted that this 1D critical point is strongly different from the O(3) QCP one due to the extreme spatial anisotropy of spin correlations. At the critical point, which corresponds to the limit of decoupled spin chains, excitations propagate only along chains. This feature, combined with unique properties of the spin-1/2 Heisenberg chain, imbues the ordered phase of weakly coupled spin-1/2 chains with the spinon confinement physics which is absent in the spatially isotropic magnetically ordered phase with spontaneously broken O(3) symmetry.

	To study the excitation spectrum of the quasi-1D spin system, 
	we utilize quantum Monte Carlo (QMC) simulations and stochastic analytic continuation (SAC) ~\cite{Computational2010Sandvik,Quantum2002Sandvik,Generalized2005Alet} to compute the spectral information of weakly coupled Heisenberg spin-1/2 chains. The predicted amplitude modes are directly observed in the numerics as the interchain interaction is reduced towards zero, and the dispersions of all low-energy modes agree nicely with analytic predictions. More importantly, we find that the amplitude mode in quasi-1D systems exhibits two novel features. First, in contrast with higher-dimensional magnets,
	the amplitude mode in quasi-1D systems is characterized by a spectral weight significantly stronger than the continuum, making it highly visible and easy to detect. Second, we find 
	that a quasi-1D spin-1/2 magnet contains {\em three}, 
	instead of two, low-energy modes. In addition to the phase and amplitude modes, visible in the dynamic spin correlation functions, an additional scalar mode emerges in the dynamic {\em bond correlation} function. Similar to the amplitude mode, this scalar mode is singletlike but exhibits different frequency and momentum dependence. 
	
	In higher dimensions, it has been known that the scalar susceptibility serves as a great tool for probing fluctuations in the singlet channel~\cite{Podolsky2011} and has been widely used in numerical studies of dimerized antiferromagnets~\cite{Dynamical2015Lohofer,Amplitude2017Qin,Lohoefer2017}. Inside the ordered SSB phase scalar fluctuations overlap with the amplitude ones but with much suppressed damping, and the scalar susceptibility exhibits a sharp peak at the amplitude mode frequency \cite{Gazit2013}. The quasi-1D limit is different. We show that in contrast to the amplitude mode which corresponds to the ``second breather" in the effective sine-Gordon description of the ordered quasi-1D magnet, the scalar mode is represented by the ``first breather", an excitation with smaller frequency which is probed via the dynamic bond-bond correlation function.

	\begin{figure*}[htp]
		\centering
		\includegraphics[width=.8\textwidth]{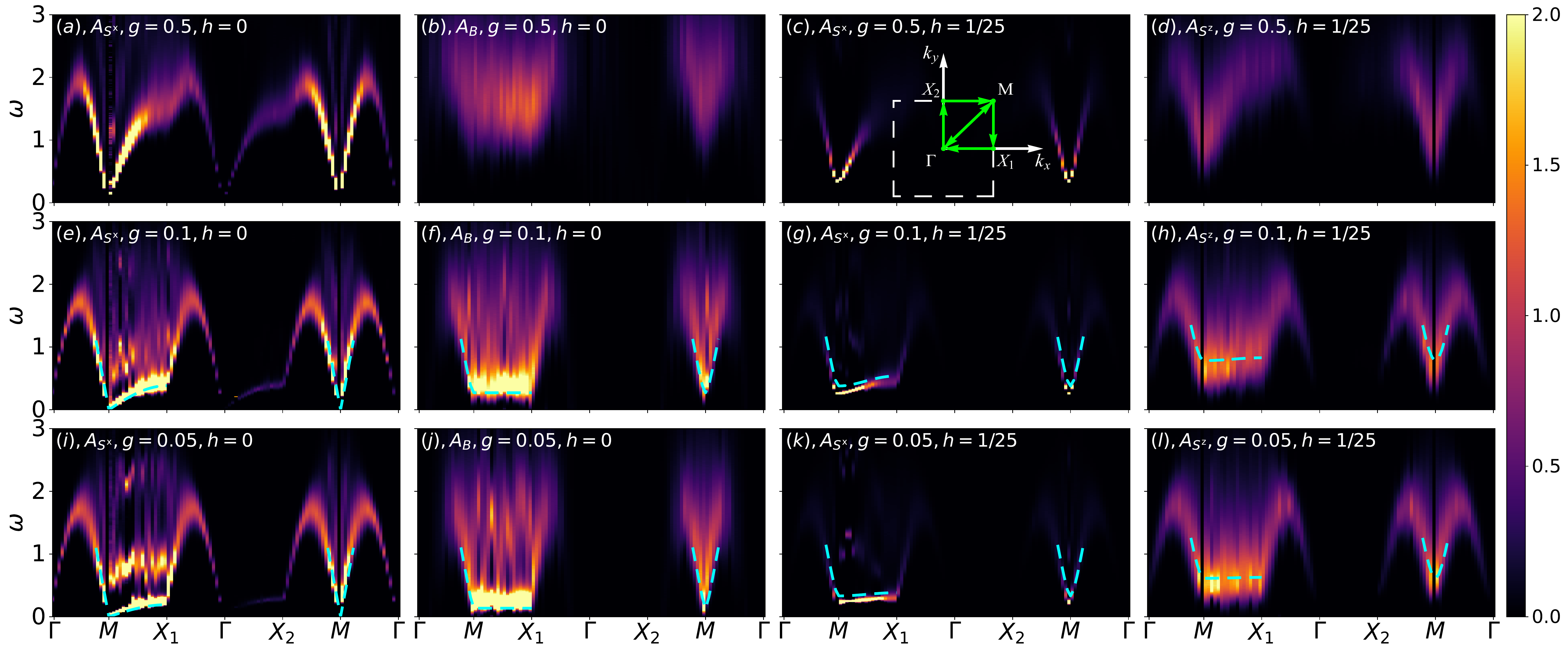}
		\caption{Spectral functions obtained from QMC-SAC. (a), (e), (i) and (b), (f), (j) show the spectra function of spin and bond operators respectively, $A_{S^{x}}(\mathbf{q},\omega)$ and $A_{B}(\mathbf{q},\omega)$, without the field, $h=0$, at different values of $g=J/J_\perp$, with the system size is $L=36$ and  inverse temperature $\beta=4L$. The dashed cyan curves in (e), (i) and (f), (j) are analytical dispersions in Eqs.~\eqref{eq:sx:dispersion} and ~\eqref{eq:sb:dispersion} with $b_h=0$.
			(c), (g), (k) and (d), (h), (l) show the phase mode spectra $A_{S^{x}}(\mathbf{q},\omega)$ and amplitude mode spectra $A_{S^{z}}(\mathbf{q},\omega)$ measured in the presence of a weak staggered field $h=1/25$, with system size $L=36$ and inverse temperature $\beta=4L$. The dashed cyan curves in (g), (k) and (h), (l) are analytical dispersions in Eqs.~\eqref{eq:sx:dispersion} and ~\eqref{eq:sz:dispersion} with finite $b_h$.}	
		\label{fig:fig2}
	\end{figure*}

	{\noindent\it The model and the QMC method.-----}
	The geometry of the problem is shown in Fig.~\ref{fig:fig1}. The Hamiltonian reads
	\begin{equation}
		H=J\sum_{\langle i,j \rangle_{\mathrm{x}}}\mathbf{S}_{i}\cdot \mathbf{S}_{j}+J_{\perp}\sum_{\langle i,j \rangle_{\mathrm{y}}}\mathbf{S}_{i}\cdot\mathbf{S}_{j} - h\sum_{i} (-1)^{i}S^{z}_{i},
		\label{eq:eq1}
	\end{equation} 
	where $\mathbf{S}_{i}=(S^{x}_{i},S^{y}_{i},S^{z}_i)$ denotes the spin-$1/2$ operator on site $i$ and $J$ ($J_{\perp}$) is the nearest-neighbor Heisenberg exchange along the $x$ ($y$) direction. We set $J=1$ and introduce ratio $g=J_{\perp}/J$ to control the crossover from decoupled 1D chains, $g=0$, to the isotropic 2D square lattice, $g=1$. The last term represents the staggered pinning field $h$, which explicitly breaks the spin-rotational symmetry. 
	
	In our QMC simulations the following three correlation functions  are measured: the transverse spin correlation function
	$G_{S^{x}}(\mathbf{q},\tau)=\frac{1}{L^2}\sum_{i,j}e^{-i\mathbf{q}\cdot(\mathbf{r}_{i}-\mathbf{r}_{j})}\langle S^{\mathrm{x}}_{i}(\tau) S^{\mathrm{x}}_{j}(0)\rangle$,
	the similarly defined longitudinal $S^z$ correlation function $G_{S^{z}}(\mathbf{q},\tau)$, and the bond correlation
	$G_{B}(\mathbf{q},\tau)=\frac{1}{L^2}\sum_{i,j}e^{-i\mathbf{q}\cdot(\mathbf{r}_{i}-\mathbf{r}_{j})}\langle B_{j}(\tau) B_{i}(0) \rangle$.
	Here $B_{i}=\mathbf{S}_{i}\cdot\mathbf{S}_{i+\hat{x}}$ is a spin singlet bond operator (dimerization order parameter) defined on a nearest-neighbor bond of the spin chain,
	$L$ is the linear system size and $\tau\in[0,\beta]$ is the imaginary time.
	In the ordered the SSB ground state with finite $\langle S^z\rangle \neq 0$ phase fluctuations (spin waves) are probed by $G_{S^{x}}$,  
	$G_{S^{z}}$ measures the amplitude fluctuations, and 
	the scalar correlation function $G_{B}$ probes correlations between bonds (energy density) ~\cite{Dynamical2015Lohofer,Amplitude2017Qin,Lohoefer2017}.
	
	It is important to notice that SSB ground state is not possible in the QMC simulation on a finite $L\times L$ system and at finite inverse temperature $\beta$.
	Therefore, in the QMC with $h=0$, there is no distinction between the phase and amplitude correlation functions, 
	$G_{S^{x}}(\mathbf{q},\tau) = G_{S^{z}}(\mathbf{q},\tau)$. Finite $h \neq 0$ breaks spin-rotational symmetry and allows one to probe the amplitude mode by measuring $G_{S^{z}}$. It also induces the $h$-dependent gap in the phase mode in $G_{S^{x}}$ \cite{ao1999,furusaki2003,essler2009}. 
	
	In order to access real-time quantum dynamics and obtain the real-frequency spectral function $A(\mathbf{q},\omega)$ from the imaginary-time correlation $G(\mathbf{q},\tau)$, $G(\textbf{q},\tau)=\frac{1}{\pi}\int_{0}^{\infty}d\omega \ A(\textbf{q},\omega) \ (e^{-\tau\omega}+e^{-(\beta-\tau)\omega})$, we employ the stochastic analytic continuation (SAC). 
	This technique, details of which are described in Refs.~\cite{Sandvik1998,Beach2004,Constrained2016Sandvik,sandvik1992generalization} and the Supplemental Material (SM)~\cite{suppl}, has been successfully applied to a broad range of quantum magnets ~\cite{Nearly2017Shao,Spin2019yining,Dynamical2018Sun,NSMA2018,CJHuang2018,NVMa2019,ZhengYan2020,YCWang2020,BKT2020,YCWang2021}. 
	
	
	{\noindent\it Analytical theory.-----} At small $g=J_\perp/J \ll 1$, a variety of exact (Bethe ansatz) and nonperturbative approaches (bosonization and renormalization group) are available. 
	In the $g=0$ limit elementary excitations of the spin chain are right- and left-moving spinons, neutral spin-1/2 fermions $\psi_{R/L, s}$, which encode an extended SU(2)$_R \times$ SU(2)$_L$ symmetry of chiral rotations at low energies. The staggered part of the lattice spin operator is expressed via spinons as $S^a_i \sim (-1)^i \psi^\dagger_{R s} \sigma^a_{s s'} \psi_{L s'} + \rm{H.c.}$, where $\sigma^a$ is the Pauli matrix. The singlet bond operator is staggered as well $B_i \sim (-1)^i \psi^\dagger_{R s} \psi_{L s} + \rm{H.c.}$. These expressions define physical response functions $G_{S^a}$ and $G_B$ of the chain. When continued to the real frequency, the response is given by the triplet and singlet spinon continua, correspondingly.
	
	Interchain interaction, $g\neq 0$, causes {\em confinement of spinons}, binding them in triplet and singlet pairs. This is easiest seen with the help of the chain mean-field theory 
	\cite{schulz1996,Quasi-one-dimensional1997Essler,sandvik1999,essler-konik} which maps the problem to the 1D sine-Gordon model by approximating the $J_\perp$ term in Eq.~\eqref{eq:eq1}
	by the interchain staggered field $2 J_\perp m_0 \sum_i (-1)^i S^z_i$ with the self-consistently determined staggered magnetic order $m_0 = (-1)^i \langle S^z_i\rangle$ along the $z$ axis \cite{suppl}. This mean field breaks spin rotational symmetry of the problem (with $h=0$). The excitation spectrum of the sine-Gordon model consists of solitons and antisolitons of mass $\Delta_0$, which describe transverse spin excitations, and their bound states, breathers. The amplitude mode, which within the low-energy mapping to the sine-Gordon model is represented by $S^z \sim \cos(\Phi/2)$, is described by the {\em second} breather, of mass $\sqrt{3}\Delta_0$. The singlet mode, which is represented as $B \sim \sin(\Phi/2)$, is instead described by the {\em first} breather, of mass $\Delta_0$, see Ref.\cite{suppl} and Refs. \cite{Quasi-one-dimensional1997Essler,essler-konik}. This brief description shows that in the system of weakly coupled spin-1/2 chains the amplitude, $S^z$, and the scalar, $B$, modes are distinct and {\em independent} excitations.
	
	Detailed calculation of spin and bond susceptibilities  
	are presented in the SM~\cite{suppl}. The dispersions of the phase ($S^x$ and $S^y$), amplitude ($S^z$) and bond ($B$) modes near $k_x=\pi$ are 
	\begin{align}
		\omega_{S^x}&=\omega_{S^y}= \Delta_0 \sqrt{1+ b_h + \cos k_y + \frac{v^2 (k_x-\pi)^2}{\Delta_0^2}}
		\label{eq:sx:dispersion}
		\\
		\omega_{S^z}&= \Delta_0 \sqrt{3(1+ b_h) + \frac{Z_2}{Z_1}\cos k_y+ \frac{v^2 (k_x-\pi)^2}{\Delta_0^2}}
		\label{eq:sz:dispersion}
		\\
		\omega_{B}&= \Delta_0\sqrt{1+b_h + \frac{v^2 (k_x-\pi)^2}{\Delta_0^2}}
		\label{eq:sb:dispersion}
	\end{align}
	Here $k_x$ ($k_y$) is the momentum along the chain (transverse to the chain) and 
	$b_h$ is a dimensionless parameter describing the the effect of the external staggered field $h$, Eq.~(\bl{S25}). At $h=0$, $b_h$ vanishes and our equations for $\omega_{S^x}$ and $\omega_{S^z}$ recover the corresponding formulae in Ref.~\cite{essler-konik}.  The velocity $v$ is $\pi J /2$ and the ratio $Z_2/Z_1\approx 0.491309$.
	
	Note that in addition to having a different mass, the dispersion of the bond mode is different from the amplitude one as well. It propagates along the chain 
	with the same velocity $v$ as spin fluctuations but is essentially dispersionless in the transverse $k_y$ direction, see Ref.\cite{suppl}. 
	
	{\noindent\it Numerical results.-----} In Fig.~\ref{fig:fig2}, we present numerical results of spectral functions for spin-spin and bond-bond correlations, with and without the pinning field $h$ and compare them with the dispersions (cyan lines) obtained from analytic theory Eqs.\eqref{eq:sx:dispersion}-\eqref{eq:sb:dispersion}. From the top to bottom row, the values of $g$ are 0.5, 0.1 and 0.05, reflecting the dimensional crossover from 2D to quasi-1D. The system has periodic boundary condition $L\times L$ with $L=36$. The QMC calculations are carried out at inverse temperature $\beta=4L$. The spectra are plotted along the high-symmetry path  
	indicated in the BZ in panel Fig.~\ref{fig:fig2}(c). The first (last) two columns of  Fig.~\ref{fig:fig2} are measured in the absence (presence) of the staggered field $h$.
	
	Key differences between the 2D [$g=0.5$, Figs.\ref{fig:fig2}(a)-\ref{fig:fig2}(d)] and quasi-1D regimes [$g=0.1$ for Figs.\ref{fig:fig2}(e)-\ref{fig:fig2}(h) and $g=0.05$ for Figs.\ref{fig:fig2}(j)-\ref{fig:fig2}(l)] are easily seen.
	For $g=0.5$ the phase mode is clearly visible in panels (a) and (c) while the amplitude and scalar fluctuations (d) and (b) exhibit only an overdamped multimagnon continuum without any sharp modes, as expected~\cite{Raman1992Canali,Longitudinal1992Ian}. As the system moves towards 1D ($g=0.1$ and $0.05$), the single magnon mode remains sharp and becomes more 1D-like (i.e. less dispersive along the interchain $M-X_1$ direction). At the same time, the spectral weight in the bond [Figs.\ref{fig:fig2}(f) and \ref{fig:fig2}(j)] and amplitude [Figs.\ref{fig:fig2}(h) and \ref{fig:fig2}(i)] sectors shifts down in energy, resulting in the emergence of the two low energy peaks in corresponding spectral densities.
	
	Let us investigate these differences closer. The first column in Fig.~\ref{fig:fig2} shows $A_{S^{x}}$ at $h=0$. Note that simulations in finite size and temperature system are necessarily done in the symmetric phase with three components of spin susceptibility degenerate $A_{S^{x}}=A_{S^{y}}=A_{S^{z}}$. The minimal spin excitation energy, measured at the $M$ point $(\pi,\pi)$, is small but finite. At $g=0.05$ it is about $0.004$. The dispersion of the lowest energy branch is well described by the pole of the RPA susceptibility Eq.~(\bl{S35}). Notice that in this magnetically disordered phase the gap is $\Delta^2 - 2 Z_1 J_\perp > 0$, as discussed above. It vanishes only in the thermodynamic limit $L=\infty$ when the SSB takes place and the spin rotational symmetry gets broken, resulting in different dispersion relations for transverse, Eq.~\eqref{eq:sx:dispersion}, and longitudinal, Eq.~\eqref{eq:sz:dispersion}, modes (with $b_h=0$).
	
	We also observe noticeable spectral intensity at higher energy, $\omega \approx 0.6-0.7$, in Figs.\ref{fig:fig2} (e) and \ref{fig:fig2}(i). We assign this to the second breather of the sine-Gordon+RPA theory, Eq.~(\bl{S36}), with the mass $\sqrt{3} \Delta$ \cite{suppl}. 
	Naturally, this feature is absent in the 2D limit, Fig.~\ref{fig:fig2}(a), where our quasi-1D arguments do not apply. This interpretation is further supported by the data for bond spectral function $A_B$, presented in the second column of Fig.~\ref{fig:fig2}. Here, one observes pronounced difference between the 2D, $g=0.5$, and 1D limits, $g=0.05$ and $0.1$: the broad and overdamped multiparticle continuum evolves into a very structured one with a sharp particlelike peak at the lowest energy for small-$g$ cases, in Figs.\ref{fig:fig2}(f) and \ref{fig:fig2}(j). This is the first breather of the sine-Gordon model, describing the scalar bond (staggered dimerization) mode, with mass $\Delta$, of weakly coupled spin chains, as described below Eq.~\eqref{eq:sb:dispersion}. As explained in the SM~\cite{suppl}, its dispersion along $k_y$ is negligible while that along $k_x$ matches Eq.~\eqref{eq:sb:dispersion} (with $b_h=0$) very well.
	
	Taken together, our data lend strong support to the description of the spin system in terms of {\em confined} spinon pairs. The spin susceptibility is described by the triplet of bound spinons and its internal excited state (the second breather) while the scalar susceptibility is represented by bound singlet pairs of spinons.
	
	To differentiate between the transverse and longitudinal fluctuations we next turn on the staggered field $h\neq 0$ along the $z$ axis. The corresponding QMC data are represented by 
	the last two columns of Fig.~\ref{fig:fig2}. Now $A_{S^{x}}=A_{S^{y}}$ measures the phase fluctuations of the order parameter (the third column), which are gapped stronger by the finite $h$, while $A_{S^{z}}$ gives the amplitude fluctuations (the last column). 
	
	\begin{figure}[htp!]
		\centering
		\includegraphics[width=.65\columnwidth]{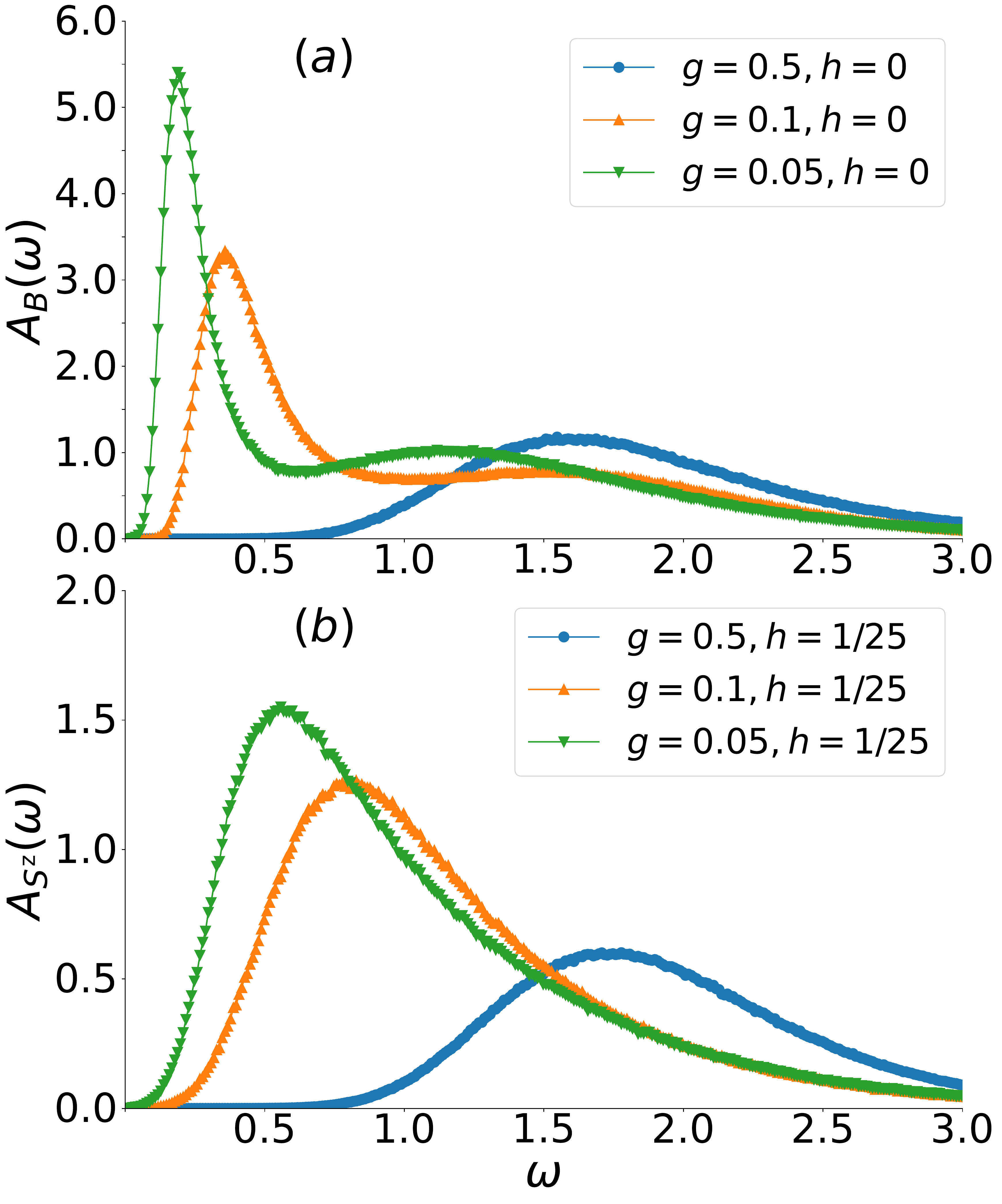}
		\caption{Frequency dependence of the spectral functions at $\mathbf{k}=(\pi,\pi/2)$ for (a) the bond-bond correlation at $h=0$ and (b) the amplitude mode at $h=1/25$.
		}
		\label{fig:fig3}
	\end{figure}
	
	\begin{figure}[htp!]
		\centering
		\includegraphics[width=.675\columnwidth]{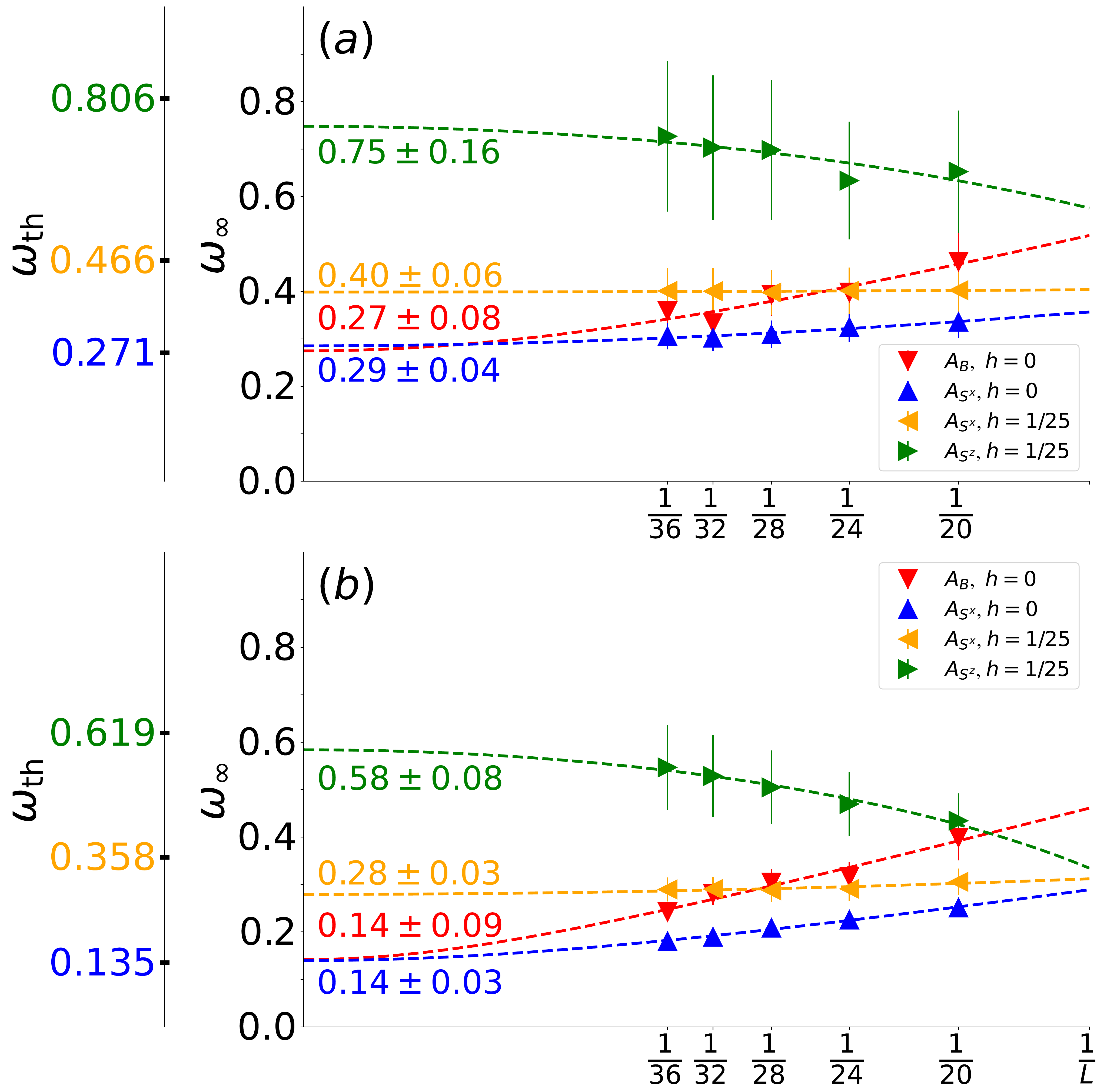}
		\caption{Finite-size analysis for  (a) $g=0.1$ and (b) $g=0.05$ at the momentum point $\mathbf{k}=(\pi,\pi/2)$. 
			The vertical $\omega_\infty$ axis shows extrapolation of the numerical data to the $L=\infty$ limit.
			Values on the left vertical $\omega_{\rm th}$ axis mark analytical predictions for the peak frequencies of different modes. See main text for details.
		}
		\label{fig:fig4}
	\end{figure}
	
	To illustrate the emergence of the amplitude and scalar modes we plot in Fig.~\ref{fig:fig3} the frequency dependence of these two spectra at different values of $g$ at the wave vector $\mathbf{k}=(\pi,\pi/2)$. In 2D ($g=0.5$), both spectra exhibit a continuum background from multimagnon excitations. As $g$ gets smaller, a peak emerges in the spectral function and becomes sharper as $g$ becomes smaller. It is seen that for the same value of $g$ the peak in $A_B$ is more narrow and occurs at a lower frequency than that in 
	$A_{S^z}$. 
	The larger linewidth of the amplitude mode is due to the stronger damping it experiences due to decays into the low energy phase fluctuations, in comparison with the bond correlation function~\cite{Podolsky2011}, while the peak's maxima difference is a unique property of quasi-1D system as Eqs. 
	\eqref{eq:sz:dispersion} and \eqref{eq:sb:dispersion} show.
	
	In Fig.~\ref{fig:fig4}, we present the finite-size analysis and extrapolate the peak frequency of each mode to the thermodynamic limit. 
	Here, we focus on the momentum point $\mathbf{k}=(\pi,\pi/2)$, at which the interchain dispersion vanishes within the RPA approximation, and the frequency of a mode is obtained by fitting the correlation function to an exponential function of the imaginary time  $\propto e^{-\omega \tau}$ (several representative cases of such fitting are presented in SM~\cite{suppl}). 
	Given the square-root form of dispersions \eqref{eq:sx:dispersion}, \eqref{eq:sz:dispersion} and \eqref{eq:sb:dispersion}, we take the following functional form for the extrapolation to infinite size $\omega_L = \sqrt{ \omega_{\infty}^2 + L_0^2/L^2}$, where $\omega_{\infty}$ and $L_0$ are fitting parameters.  The results for so obtained $\omega_{\infty}$ are presented
	to the right of the $\omega_{\infty}$ axis in Fig.~\ref{fig:fig4}. As discussed above, without the staggered field  the lowest energy peak in $A_{S^x} (h=0)$ describes coherent threefold degenerate mode $\omega_{S^x}$. At finite $h=1/25$, the degeneracy is removed and $\omega_{S^x}$ and $\omega_{S^z}$ scale to different limits. Within the sine-Gordon description $\omega_{S^z}/\omega_{S^x} = \sqrt{3}$, see Eqs.~\eqref{eq:sx:dispersion} and \eqref{eq:sz:dispersion}. Figure~\ref{fig:fig4} shows that this ratio extrapolates to $1.8$ for $g=0.1$ and to $2.0$ for $g=0.05$ at $L=\infty$. The $\omega_{\rm th}$ axis in Fig.~\ref{fig:fig4} shows analytical predictions for $\omega_{S^x}, \omega_{S^z}$, which are calculated as functions of $g$ and $h$, without any adjustable parameters, in the SM~\cite{suppl}. This parameter-free comparison is seen to work reasonably well. In addition, in agreement with analytical predictions, Fig.~\ref{fig:fig4} shows that for $h=0$, $\omega_{S^x}$ and $\omega_B$ extrapolate to the same limit, just as Eqs. 
	\eqref{eq:sx:dispersion} and \eqref{eq:sb:dispersion} with $b_h=0$ require.
	
	{\noindent\it Summary.---} Our main finding is that in quasi-1D limit collective amplitude and scalar modes appear in dynamic spin and bond susceptibilities, as is seen in Figs.~\ref{fig:fig2} and~~\ref{fig:fig3}, while contributions from multi-particle continua are much weaker and subleading. This needs to be contrasted with the vicinity of the O(3) QCP in higher dimensions~\cite{Gazit2013,Amplitude2017Qin,Lohoefer2017}, where the situation is the opposite. 
	This makes spin-1/2 quasi-1D magnets an attractive candidate systems for studying collective quantum dynamics theoretically and experimentally.
	
	\begin{acknowledgements} 
		O.A.S. thanks Sasha Chernyshev for insightful discussions. C.K.Z., Z.Y. and Z.Y.M. acknowledge the support from the RGC of Hong Kong SAR China (Grants No. 17303019, No.17301420, and No.AoE/P-701/20), and MOST through the National Key Research and Development Program (2016YFA0300502). 
		O.A.S. was supported by the NSF CMMT program under Grant No.DMR-1928919.
		H.Q.W. is thankful for supports from NSFC-11804401 and the Fundamental Research Funds for the Central Universities (19lgpy266).
		We thank the Computational Initiative at the Faculty of Science and the Information Technology Services at the University of Hong Kong and the National Supercomputer Centers in Guangzhou and Beijng PARATERA Tech CO.,Ltd. for their technical support and providing generous HPC resources that have contributed to the research results reported within this paper. This research was initiated at the Aspen Center for Physics, supported by NSF PHY-1066293.
	\end{acknowledgements}
	
	\bibliography{ref}

\begin{thebibliography}{45}%
\makeatletter
\providecommand \@ifxundefined [1]{%
 \@ifx{#1\undefined}
}%
\providecommand \@ifnum [1]{%
 \ifnum #1\expandafter \@firstoftwo
 \else \expandafter \@secondoftwo
 \fi
}%
\providecommand \@ifx [1]{%
 \ifx #1\expandafter \@firstoftwo
 \else \expandafter \@secondoftwo
 \fi
}%
\providecommand \natexlab [1]{#1}%
\providecommand \enquote  [1]{``#1''}%
\providecommand \bibnamefont  [1]{#1}%
\providecommand \bibfnamefont [1]{#1}%
\providecommand \citenamefont [1]{#1}%
\providecommand \href@noop [0]{\@secondoftwo}%
\providecommand \href [0]{\begingroup \@sanitize@url \@href}%
\providecommand \@href[1]{\@@startlink{#1}\@@href}%
\providecommand \@@href[1]{\endgroup#1\@@endlink}%
\providecommand \@sanitize@url [0]{\catcode `\\12\catcode `\$12\catcode
  `\&12\catcode `\#12\catcode `\^12\catcode `\_12\catcode `\%12\relax}%
\providecommand \@@startlink[1]{}%
\providecommand \@@endlink[0]{}%
\providecommand \url  [0]{\begingroup\@sanitize@url \@url }%
\providecommand \@url [1]{\endgroup\@href {#1}{\urlprefix }}%
\providecommand \urlprefix  [0]{URL }%
\providecommand \Eprint [0]{\href }%
\providecommand \doibase [0]{http://dx.doi.org/}%
\providecommand \selectlanguage [0]{\@gobble}%
\providecommand \bibinfo  [0]{\@secondoftwo}%
\providecommand \bibfield  [0]{\@secondoftwo}%
\providecommand \translation [1]{[#1]}%
\providecommand \BibitemOpen [0]{}%
\providecommand \bibitemStop [0]{}%
\providecommand \bibitemNoStop [0]{.\EOS\space}%
\providecommand \EOS [0]{\spacefactor3000\relax}%
\providecommand \BibitemShut  [1]{\csname bibitem#1\endcsname}%
\let\auto@bib@innerbib\@empty
\bibitem [{\citenamefont {Merchant}\ \emph {et~al.}(2014)\citenamefont
  {Merchant}, \citenamefont {Normand}, \citenamefont {Kr\"amer}, \citenamefont
  {Boehm}, \citenamefont {McMorrow},\ and\ \citenamefont
  {R\"uegg}}]{Merchant2014}%
  \BibitemOpen
  \bibfield  {author} {\bibinfo {author} {\bibfnamefont {P.}~\bibnamefont
  {Merchant}}, \bibinfo {author} {\bibfnamefont {B.}~\bibnamefont {Normand}},
  \bibinfo {author} {\bibfnamefont {K.~W.}\ \bibnamefont {Kr\"amer}}, \bibinfo
  {author} {\bibfnamefont {M.}~\bibnamefont {Boehm}}, \bibinfo {author}
  {\bibfnamefont {D.~F.}\ \bibnamefont {McMorrow}}, \ and\ \bibinfo {author}
  {\bibfnamefont {C.}~\bibnamefont {R\"uegg}},\ }\href {\doibase
  10.1038/nphys2902} {\bibfield  {journal} {\bibinfo  {journal} {Nature
  Physics}\ }\textbf {\bibinfo {volume} {10}},\ \bibinfo {pages} {373 }
  (\bibinfo {year} {2014})}\BibitemShut {NoStop}%
\bibitem [{\citenamefont {Lake}\ \emph {et~al.}(2000)\citenamefont {Lake},
  \citenamefont {Tennant},\ and\ \citenamefont {Nagler}}]{Novel2000Lake}%
  \BibitemOpen
  \bibfield  {author} {\bibinfo {author} {\bibfnamefont {B.}~\bibnamefont
  {Lake}}, \bibinfo {author} {\bibfnamefont {D.~A.}\ \bibnamefont {Tennant}}, \
  and\ \bibinfo {author} {\bibfnamefont {S.~E.}\ \bibnamefont {Nagler}},\
  }\href {\doibase 10.1103/PhysRevLett.85.832} {\bibfield  {journal} {\bibinfo
  {journal} {Phys. Rev. Lett.}\ }\textbf {\bibinfo {volume} {85}},\ \bibinfo
  {pages} {832} (\bibinfo {year} {2000})}\BibitemShut {NoStop}%
\bibitem [{\citenamefont {Lake}\ \emph {et~al.}(2005)\citenamefont {Lake},
  \citenamefont {Tennant},\ and\ \citenamefont
  {Nagler}}]{Longitudinal2005Lake}%
  \BibitemOpen
  \bibfield  {author} {\bibinfo {author} {\bibfnamefont {B.}~\bibnamefont
  {Lake}}, \bibinfo {author} {\bibfnamefont {D.~A.}\ \bibnamefont {Tennant}}, \
  and\ \bibinfo {author} {\bibfnamefont {S.~E.}\ \bibnamefont {Nagler}},\
  }\href {\doibase 10.1103/PhysRevB.71.134412} {\bibfield  {journal} {\bibinfo
  {journal} {Phys. Rev. B}\ }\textbf {\bibinfo {volume} {71}},\ \bibinfo
  {pages} {134412} (\bibinfo {year} {2005})}\BibitemShut {NoStop}%
\bibitem [{\citenamefont {Zheludev}\ \emph {et~al.}(2001)\citenamefont
  {Zheludev}, \citenamefont {Kenzelmann}, \citenamefont {Raymond},
  \citenamefont {Masuda}, \citenamefont {Uchinokura},\ and\ \citenamefont
  {Lee}}]{Spin2001Zheludev}%
  \BibitemOpen
  \bibfield  {author} {\bibinfo {author} {\bibfnamefont {A.}~\bibnamefont
  {Zheludev}}, \bibinfo {author} {\bibfnamefont {M.}~\bibnamefont
  {Kenzelmann}}, \bibinfo {author} {\bibfnamefont {S.}~\bibnamefont {Raymond}},
  \bibinfo {author} {\bibfnamefont {T.}~\bibnamefont {Masuda}}, \bibinfo
  {author} {\bibfnamefont {K.}~\bibnamefont {Uchinokura}}, \ and\ \bibinfo
  {author} {\bibfnamefont {S.-H.}\ \bibnamefont {Lee}},\ }\href {\doibase
  10.1103/PhysRevB.65.014402} {\bibfield  {journal} {\bibinfo  {journal} {Phys.
  Rev. B}\ }\textbf {\bibinfo {volume} {65}},\ \bibinfo {pages} {014402}
  (\bibinfo {year} {2001})}\BibitemShut {NoStop}%
\bibitem [{\citenamefont {Bera}\ \emph {et~al.}(2017)\citenamefont {Bera},
  \citenamefont {Lake}, \citenamefont {Essler}, \citenamefont {Vanderstraeten},
  \citenamefont {Hubig}, \citenamefont {Schollw\"ock}, \citenamefont {Islam},
  \citenamefont {Schneidewind},\ and\ \citenamefont
  {Quintero-Castro}}]{Bera2017}%
  \BibitemOpen
  \bibfield  {author} {\bibinfo {author} {\bibfnamefont {A.~K.}\ \bibnamefont
  {Bera}}, \bibinfo {author} {\bibfnamefont {B.}~\bibnamefont {Lake}}, \bibinfo
  {author} {\bibfnamefont {F.~H.~L.}\ \bibnamefont {Essler}}, \bibinfo {author}
  {\bibfnamefont {L.}~\bibnamefont {Vanderstraeten}}, \bibinfo {author}
  {\bibfnamefont {C.}~\bibnamefont {Hubig}}, \bibinfo {author} {\bibfnamefont
  {U.}~\bibnamefont {Schollw\"ock}}, \bibinfo {author} {\bibfnamefont {A.~T.
  M.~N.}\ \bibnamefont {Islam}}, \bibinfo {author} {\bibfnamefont
  {A.}~\bibnamefont {Schneidewind}}, \ and\ \bibinfo {author} {\bibfnamefont
  {D.~L.}\ \bibnamefont {Quintero-Castro}},\ }\href {\doibase
  10.1103/PhysRevB.96.054423} {\bibfield  {journal} {\bibinfo  {journal} {Phys.
  Rev. B}\ }\textbf {\bibinfo {volume} {96}},\ \bibinfo {pages} {054423}
  (\bibinfo {year} {2017})}\BibitemShut {NoStop}%
\bibitem [{\citenamefont {Gannon}\ \emph {et~al.}(2019)\citenamefont {Gannon},
  \citenamefont {Zaliznyak}, \citenamefont {Wu}, \citenamefont {Feiguin},
  \citenamefont {Tsvelik}, \citenamefont {Demmel}, \citenamefont {Qiu},
  \citenamefont {Copley}, \citenamefont {Kim},\ and\ \citenamefont
  {Aronson}}]{Gannon2019}%
  \BibitemOpen
  \bibfield  {author} {\bibinfo {author} {\bibfnamefont {W.~J.}\ \bibnamefont
  {Gannon}}, \bibinfo {author} {\bibfnamefont {I.~A.}\ \bibnamefont
  {Zaliznyak}}, \bibinfo {author} {\bibfnamefont {L.~S.}\ \bibnamefont {Wu}},
  \bibinfo {author} {\bibfnamefont {A.~E.}\ \bibnamefont {Feiguin}}, \bibinfo
  {author} {\bibfnamefont {A.~M.}\ \bibnamefont {Tsvelik}}, \bibinfo {author}
  {\bibfnamefont {F.}~\bibnamefont {Demmel}}, \bibinfo {author} {\bibfnamefont
  {Y.}~\bibnamefont {Qiu}}, \bibinfo {author} {\bibfnamefont {J.~R.~D.}\
  \bibnamefont {Copley}}, \bibinfo {author} {\bibfnamefont {M.~S.}\
  \bibnamefont {Kim}}, \ and\ \bibinfo {author} {\bibfnamefont {M.~C.}\
  \bibnamefont {Aronson}},\ }\href {\doibase 10.1038/s41467-019-08715-y}
  {\bibfield  {journal} {\bibinfo  {journal} {Nat. Commun.}\ }\textbf {\bibinfo
  {volume} {10}},\ \bibinfo {pages} {1123} (\bibinfo {year}
  {2019})}\BibitemShut {NoStop}%
\bibitem [{\citenamefont {Sherman}\ \emph {et~al.}(2015)\citenamefont
  {Sherman}, \citenamefont {Pracht}, \citenamefont {Gorshunov}, \citenamefont
  {Poran}, \citenamefont {Jesudasan}, \citenamefont {Chand}, \citenamefont
  {Raychaudhuri}, \citenamefont {Swanson}, \citenamefont {Trivedi},
  \citenamefont {Auerbach}, \citenamefont {Scheffler}, \citenamefont
  {Frydman},\ and\ \citenamefont {Dressel}}]{Sherman2015}%
  \BibitemOpen
  \bibfield  {author} {\bibinfo {author} {\bibfnamefont {D.}~\bibnamefont
  {Sherman}}, \bibinfo {author} {\bibfnamefont {U.~S.}\ \bibnamefont {Pracht}},
  \bibinfo {author} {\bibfnamefont {B.}~\bibnamefont {Gorshunov}}, \bibinfo
  {author} {\bibfnamefont {S.}~\bibnamefont {Poran}}, \bibinfo {author}
  {\bibfnamefont {J.}~\bibnamefont {Jesudasan}}, \bibinfo {author}
  {\bibfnamefont {M.}~\bibnamefont {Chand}}, \bibinfo {author} {\bibfnamefont
  {P.}~\bibnamefont {Raychaudhuri}}, \bibinfo {author} {\bibfnamefont
  {M.}~\bibnamefont {Swanson}}, \bibinfo {author} {\bibfnamefont
  {N.}~\bibnamefont {Trivedi}}, \bibinfo {author} {\bibfnamefont
  {A.}~\bibnamefont {Auerbach}}, \bibinfo {author} {\bibfnamefont
  {M.}~\bibnamefont {Scheffler}}, \bibinfo {author} {\bibfnamefont
  {A.}~\bibnamefont {Frydman}}, \ and\ \bibinfo {author} {\bibfnamefont
  {M.}~\bibnamefont {Dressel}},\ }\href {\doibase 10.1038/nphys3227} {\bibfield
   {journal} {\bibinfo  {journal} {Nature Physics}\ }\textbf {\bibinfo {volume}
  {11}},\ \bibinfo {pages} {188 } (\bibinfo {year} {2015})}\BibitemShut
  {NoStop}%
\bibitem [{\citenamefont {Shimano}\ and\ \citenamefont
  {Tsuji}(2020)}]{Shimano2020}%
  \BibitemOpen
  \bibfield  {author} {\bibinfo {author} {\bibfnamefont {R.}~\bibnamefont
  {Shimano}}\ and\ \bibinfo {author} {\bibfnamefont {N.}~\bibnamefont
  {Tsuji}},\ }\href {\doibase 10.1146/annurev-conmatphys-031119-050813}
  {\bibfield  {journal} {\bibinfo  {journal} {Annu. Rev. Condens. Matter
  Phys.}\ }\textbf {\bibinfo {volume} {11}},\ \bibinfo {pages} {103} (\bibinfo
  {year} {2020})}\BibitemShut {NoStop}%
\bibitem [{\citenamefont {Sachdev}(1999)}]{Sachdev1999}%
  \BibitemOpen
  \bibfield  {author} {\bibinfo {author} {\bibfnamefont {S.}~\bibnamefont
  {Sachdev}},\ }\href {\doibase 10.1103/PhysRevB.59.14054} {\bibfield
  {journal} {\bibinfo  {journal} {Phys. Rev. B}\ }\textbf {\bibinfo {volume}
  {59}},\ \bibinfo {pages} {14054} (\bibinfo {year} {1999})}\BibitemShut
  {NoStop}%
\bibitem [{\citenamefont {Podolsky}\ \emph {et~al.}(2011)\citenamefont
  {Podolsky}, \citenamefont {Auerbach},\ and\ \citenamefont
  {Arovas}}]{Podolsky2011}%
  \BibitemOpen
  \bibfield  {author} {\bibinfo {author} {\bibfnamefont {D.}~\bibnamefont
  {Podolsky}}, \bibinfo {author} {\bibfnamefont {A.}~\bibnamefont {Auerbach}},
  \ and\ \bibinfo {author} {\bibfnamefont {D.~P.}\ \bibnamefont {Arovas}},\
  }\href {\doibase 10.1103/PhysRevB.84.174522} {\bibfield  {journal} {\bibinfo
  {journal} {Phys. Rev. B}\ }\textbf {\bibinfo {volume} {84}},\ \bibinfo
  {pages} {174522} (\bibinfo {year} {2011})}\BibitemShut {NoStop}%
\bibitem [{\citenamefont {Gazit}\ \emph {et~al.}(2013)\citenamefont {Gazit},
  \citenamefont {Podolsky},\ and\ \citenamefont {Auerbach}}]{Gazit2013}%
  \BibitemOpen
  \bibfield  {author} {\bibinfo {author} {\bibfnamefont {S.}~\bibnamefont
  {Gazit}}, \bibinfo {author} {\bibfnamefont {D.}~\bibnamefont {Podolsky}}, \
  and\ \bibinfo {author} {\bibfnamefont {A.}~\bibnamefont {Auerbach}},\ }\href
  {\doibase 10.1103/PhysRevLett.110.140401} {\bibfield  {journal} {\bibinfo
  {journal} {Phys. Rev. Lett.}\ }\textbf {\bibinfo {volume} {110}},\ \bibinfo
  {pages} {140401} (\bibinfo {year} {2013})}\BibitemShut {NoStop}%
\bibitem [{\citenamefont {{Scammell}}\ and\ \citenamefont
  {{Sushkov}}(2017)}]{Sushkov2017}%
  \BibitemOpen
  \bibfield  {author} {\bibinfo {author} {\bibfnamefont {H.~D.}\ \bibnamefont
  {{Scammell}}}\ and\ \bibinfo {author} {\bibfnamefont {O.~P.}\ \bibnamefont
  {{Sushkov}}},\ }\href@noop {} {\bibfield  {journal} {\bibinfo  {journal}
  {arXiv e-prints}\ ,\ \bibinfo {eid} {arXiv:1705.09007}} (\bibinfo {year}
  {2017})},\ \Eprint {http://arxiv.org/abs/1705.09007} {arXiv:1705.09007
  [cond-mat.str-el]} \BibitemShut {NoStop}%
\bibitem [{\citenamefont {Canali}\ and\ \citenamefont
  {Girvin}(1992)}]{Raman1992Canali}%
  \BibitemOpen
  \bibfield  {author} {\bibinfo {author} {\bibfnamefont {C.~M.}\ \bibnamefont
  {Canali}}\ and\ \bibinfo {author} {\bibfnamefont {S.~M.}\ \bibnamefont
  {Girvin}},\ }\href {\doibase 10.1103/PhysRevB.45.7127} {\bibfield  {journal}
  {\bibinfo  {journal} {Phys. Rev. B}\ }\textbf {\bibinfo {volume} {45}},\
  \bibinfo {pages} {7127} (\bibinfo {year} {1992})}\BibitemShut {NoStop}%
\bibitem [{\citenamefont {Affleck}\ and\ \citenamefont
  {Wellman}(1992)}]{Longitudinal1992Ian}%
  \BibitemOpen
  \bibfield  {author} {\bibinfo {author} {\bibfnamefont {I.}~\bibnamefont
  {Affleck}}\ and\ \bibinfo {author} {\bibfnamefont {G.~F.}\ \bibnamefont
  {Wellman}},\ }\href {\doibase 10.1103/PhysRevB.46.8934} {\bibfield  {journal}
  {\bibinfo  {journal} {Phys. Rev. B}\ }\textbf {\bibinfo {volume} {46}},\
  \bibinfo {pages} {8934} (\bibinfo {year} {1992})}\BibitemShut {NoStop}%
\bibitem [{\citenamefont {Schulz}(1996)}]{schulz1996}%
  \BibitemOpen
  \bibfield  {author} {\bibinfo {author} {\bibfnamefont {H.~J.}\ \bibnamefont
  {Schulz}},\ }\href {\doibase 10.1103/PhysRevLett.77.2790} {\bibfield
  {journal} {\bibinfo  {journal} {Phys. Rev. Lett.}\ }\textbf {\bibinfo
  {volume} {77}},\ \bibinfo {pages} {2790} (\bibinfo {year}
  {1996})}\BibitemShut {NoStop}%
\bibitem [{\citenamefont {Essler}\ \emph {et~al.}(1997)\citenamefont {Essler},
  \citenamefont {Tsvelik},\ and\ \citenamefont
  {Delfino}}]{Quasi-one-dimensional1997Essler}%
  \BibitemOpen
  \bibfield  {author} {\bibinfo {author} {\bibfnamefont {F.~H.~L.}\
  \bibnamefont {Essler}}, \bibinfo {author} {\bibfnamefont {A.~M.}\
  \bibnamefont {Tsvelik}}, \ and\ \bibinfo {author} {\bibfnamefont
  {G.}~\bibnamefont {Delfino}},\ }\href {\doibase 10.1103/PhysRevB.56.11001}
  {\bibfield  {journal} {\bibinfo  {journal} {Phys. Rev. B}\ }\textbf {\bibinfo
  {volume} {56}},\ \bibinfo {pages} {11001} (\bibinfo {year}
  {1997})}\BibitemShut {NoStop}%
\bibitem [{\citenamefont {Loh\"ofer}\ \emph {et~al.}(2015)\citenamefont
  {Loh\"ofer}, \citenamefont {Coletta}, \citenamefont {Joshi}, \citenamefont
  {Assaad}, \citenamefont {Vojta}, \citenamefont {Wessel},\ and\ \citenamefont
  {Mila}}]{Dynamical2015Lohofer}%
  \BibitemOpen
  \bibfield  {author} {\bibinfo {author} {\bibfnamefont {M.}~\bibnamefont
  {Loh\"ofer}}, \bibinfo {author} {\bibfnamefont {T.}~\bibnamefont {Coletta}},
  \bibinfo {author} {\bibfnamefont {D.~G.}\ \bibnamefont {Joshi}}, \bibinfo
  {author} {\bibfnamefont {F.~F.}\ \bibnamefont {Assaad}}, \bibinfo {author}
  {\bibfnamefont {M.}~\bibnamefont {Vojta}}, \bibinfo {author} {\bibfnamefont
  {S.}~\bibnamefont {Wessel}}, \ and\ \bibinfo {author} {\bibfnamefont
  {F.}~\bibnamefont {Mila}},\ }\href {\doibase 10.1103/PhysRevB.92.245137}
  {\bibfield  {journal} {\bibinfo  {journal} {Phys. Rev. B}\ }\textbf {\bibinfo
  {volume} {92}},\ \bibinfo {pages} {245137} (\bibinfo {year}
  {2015})}\BibitemShut {NoStop}%
\bibitem [{\citenamefont {Qin}\ \emph {et~al.}(2017)\citenamefont {Qin},
  \citenamefont {Normand}, \citenamefont {Sandvik},\ and\ \citenamefont
  {Meng}}]{Amplitude2017Qin}%
  \BibitemOpen
  \bibfield  {author} {\bibinfo {author} {\bibfnamefont {Y.~Q.}\ \bibnamefont
  {Qin}}, \bibinfo {author} {\bibfnamefont {B.}~\bibnamefont {Normand}},
  \bibinfo {author} {\bibfnamefont {A.~W.}\ \bibnamefont {Sandvik}}, \ and\
  \bibinfo {author} {\bibfnamefont {Z.~Y.}\ \bibnamefont {Meng}},\ }\href
  {\doibase 10.1103/PhysRevLett.118.147207} {\bibfield  {journal} {\bibinfo
  {journal} {Phys. Rev. Lett.}\ }\textbf {\bibinfo {volume} {118}},\ \bibinfo
  {pages} {147207} (\bibinfo {year} {2017})}\BibitemShut {NoStop}%
\bibitem [{\citenamefont {Loh\"ofer}\ and\ \citenamefont
  {Wessel}(2017)}]{Lohoefer2017}%
  \BibitemOpen
  \bibfield  {author} {\bibinfo {author} {\bibfnamefont {M.}~\bibnamefont
  {Loh\"ofer}}\ and\ \bibinfo {author} {\bibfnamefont {S.}~\bibnamefont
  {Wessel}},\ }\href {\doibase 10.1103/PhysRevLett.118.147206} {\bibfield
  {journal} {\bibinfo  {journal} {Phys. Rev. Lett.}\ }\textbf {\bibinfo
  {volume} {118}},\ \bibinfo {pages} {147206} (\bibinfo {year}
  {2017})}\BibitemShut {NoStop}%
\bibitem [{\citenamefont {Swanson}\ \emph {et~al.}(2014)\citenamefont
  {Swanson}, \citenamefont {Loh}, \citenamefont {Randeria},\ and\ \citenamefont
  {Trivedi}}]{Swanson2014}%
  \BibitemOpen
  \bibfield  {author} {\bibinfo {author} {\bibfnamefont {M.}~\bibnamefont
  {Swanson}}, \bibinfo {author} {\bibfnamefont {Y.~L.}\ \bibnamefont {Loh}},
  \bibinfo {author} {\bibfnamefont {M.}~\bibnamefont {Randeria}}, \ and\
  \bibinfo {author} {\bibfnamefont {N.}~\bibnamefont {Trivedi}},\ }\href
  {\doibase 10.1103/PhysRevX.4.021007} {\bibfield  {journal} {\bibinfo
  {journal} {Phys. Rev. X}\ }\textbf {\bibinfo {volume} {4}},\ \bibinfo {pages}
  {021007} (\bibinfo {year} {2014})}\BibitemShut {NoStop}%
\bibitem [{\citenamefont {Sandvik}(2010)}]{Computational2010Sandvik}%
  \BibitemOpen
  \bibfield  {author} {\bibinfo {author} {\bibfnamefont {A.~W.}\ \bibnamefont
  {Sandvik}},\ }in\ \href {https://aip.scitation.org/doi/abs/10.1063/1.3518900}
  {\emph {\bibinfo {booktitle} {AIP Conference Proceedings}}},\ Vol.\ \bibinfo
  {volume} {1297}\ (\bibinfo {organization} {AIP},\ \bibinfo {year} {2010})\
  pp.\ \bibinfo {pages} {135--338}\BibitemShut {NoStop}%
\bibitem [{\citenamefont {Sylju\aa{}sen}\ and\ \citenamefont
  {Sandvik}(2002)}]{Quantum2002Sandvik}%
  \BibitemOpen
  \bibfield  {author} {\bibinfo {author} {\bibfnamefont {O.~F.}\ \bibnamefont
  {Sylju\aa{}sen}}\ and\ \bibinfo {author} {\bibfnamefont {A.~W.}\ \bibnamefont
  {Sandvik}},\ }\href {\doibase 10.1103/PhysRevE.66.046701} {\bibfield
  {journal} {\bibinfo  {journal} {Phys. Rev. E}\ }\textbf {\bibinfo {volume}
  {66}},\ \bibinfo {pages} {046701} (\bibinfo {year} {2002})}\BibitemShut
  {NoStop}%
\bibitem [{\citenamefont {Alet}\ \emph {et~al.}(2005)\citenamefont {Alet},
  \citenamefont {Wessel},\ and\ \citenamefont {Troyer}}]{Generalized2005Alet}%
  \BibitemOpen
  \bibfield  {author} {\bibinfo {author} {\bibfnamefont {F.}~\bibnamefont
  {Alet}}, \bibinfo {author} {\bibfnamefont {S.}~\bibnamefont {Wessel}}, \ and\
  \bibinfo {author} {\bibfnamefont {M.}~\bibnamefont {Troyer}},\ }\href
  {\doibase 10.1103/PhysRevE.71.036706} {\bibfield  {journal} {\bibinfo
  {journal} {Phys. Rev. E}\ }\textbf {\bibinfo {volume} {71}},\ \bibinfo
  {pages} {036706} (\bibinfo {year} {2005})}\BibitemShut {NoStop}%
\bibitem [{\citenamefont {Affleck}\ and\ \citenamefont
  {Oshikawa}(1999)}]{ao1999}%
  \BibitemOpen
  \bibfield  {author} {\bibinfo {author} {\bibfnamefont {I.}~\bibnamefont
  {Affleck}}\ and\ \bibinfo {author} {\bibfnamefont {M.}~\bibnamefont
  {Oshikawa}},\ }\href {\doibase 10.1103/PhysRevB.60.1038} {\bibfield
  {journal} {\bibinfo  {journal} {Phys. Rev. B}\ }\textbf {\bibinfo {volume}
  {60}},\ \bibinfo {pages} {1038} (\bibinfo {year} {1999})}\BibitemShut
  {NoStop}%
\bibitem [{\citenamefont {Essler}\ \emph {et~al.}(2003)\citenamefont {Essler},
  \citenamefont {Furusaki},\ and\ \citenamefont {Hikihara}}]{furusaki2003}%
  \BibitemOpen
  \bibfield  {author} {\bibinfo {author} {\bibfnamefont {F.~H.~L.}\
  \bibnamefont {Essler}}, \bibinfo {author} {\bibfnamefont {A.}~\bibnamefont
  {Furusaki}}, \ and\ \bibinfo {author} {\bibfnamefont {T.}~\bibnamefont
  {Hikihara}},\ }\href {\doibase 10.1103/PhysRevB.68.064410} {\bibfield
  {journal} {\bibinfo  {journal} {Phys. Rev. B}\ }\textbf {\bibinfo {volume}
  {68}},\ \bibinfo {pages} {064410} (\bibinfo {year} {2003})}\BibitemShut
  {NoStop}%
\bibitem [{\citenamefont {Kuzmenko}\ and\ \citenamefont
  {Essler}(2009)}]{essler2009}%
  \BibitemOpen
  \bibfield  {author} {\bibinfo {author} {\bibfnamefont {I.}~\bibnamefont
  {Kuzmenko}}\ and\ \bibinfo {author} {\bibfnamefont {F.~H.~L.}\ \bibnamefont
  {Essler}},\ }\href {\doibase 10.1103/PhysRevB.79.024402} {\bibfield
  {journal} {\bibinfo  {journal} {Phys. Rev. B}\ }\textbf {\bibinfo {volume}
  {79}},\ \bibinfo {pages} {024402} (\bibinfo {year} {2009})}\BibitemShut
  {NoStop}%
\bibitem [{\citenamefont {Sandvik}(1998)}]{Sandvik1998}%
  \BibitemOpen
  \bibfield  {author} {\bibinfo {author} {\bibfnamefont {A.~W.}\ \bibnamefont
  {Sandvik}},\ }\href {\doibase 10.1103/PhysRevB.57.10287} {\bibfield
  {journal} {\bibinfo  {journal} {Phys. Rev. B}\ }\textbf {\bibinfo {volume}
  {57}},\ \bibinfo {pages} {10287} (\bibinfo {year} {1998})}\BibitemShut
  {NoStop}%
\bibitem [{\citenamefont {{Beach}}(2004)}]{Beach2004}%
  \BibitemOpen
  \bibfield  {author} {\bibinfo {author} {\bibfnamefont {K.~S.~D.}\
  \bibnamefont {{Beach}}},\ }\href@noop {} {\bibfield  {journal} {\bibinfo
  {journal} {arXiv e-prints}\ ,\ \bibinfo {eid} {cond-mat/0403055}} (\bibinfo
  {year} {2004})},\ \Eprint {http://arxiv.org/abs/cond-mat/0403055}
  {arXiv:cond-mat/0403055 [cond-mat.str-el]} \BibitemShut {NoStop}%
\bibitem [{\citenamefont {Sandvik}(2016)}]{Constrained2016Sandvik}%
  \BibitemOpen
  \bibfield  {author} {\bibinfo {author} {\bibfnamefont {A.~W.}\ \bibnamefont
  {Sandvik}},\ }\href {\doibase 10.1103/PhysRevE.94.063308} {\bibfield
  {journal} {\bibinfo  {journal} {Phys. Rev. E}\ }\textbf {\bibinfo {volume}
  {94}},\ \bibinfo {pages} {063308} (\bibinfo {year} {2016})}\BibitemShut
  {NoStop}%
\bibitem [{\citenamefont {Sandvik}(1992)}]{sandvik1992generalization}%
  \BibitemOpen
  \bibfield  {author} {\bibinfo {author} {\bibfnamefont {A.~W.}\ \bibnamefont
  {Sandvik}},\ }\href
  {https://iopscience-iop-org.eproxy.lib.hku.hk/article/10.1088/0305-4470/25/13/017/meta}
  {\bibfield  {journal} {\bibinfo  {journal} {J. Phys. A Math. Theor.}\
  }\textbf {\bibinfo {volume} {25}},\ \bibinfo {pages} {3667} (\bibinfo {year}
  {1992})}\BibitemShut {NoStop}%
\bibitem [{sup()}]{suppl}%
  \BibitemOpen
  \href@noop {} {\bibinfo  {journal} {See the Supplemental Material for details
  of the QMC-SAC implementation of the quasi-1D quantum magnet, the QMC
  measurement of dynamic bond correlation function and the detailed discussion
  on the analytical calculation of the excitation spectra, which includes
  Refs.~\cite{Nearly2017Shao,Constrained2016Sandvik,Amplitude2017Qin,Computational2010Sandvik,Quantum2002Sandvik,Generalized2005Alet,Lohoefer2017,essler-konik,ao1999,Lukyanov1997,sandvik1999,hikihara2010,Quasi-one-dimensional1997Essler}}\
  }\BibitemShut {NoStop}%
\bibitem [{\citenamefont {Shao}\ \emph {et~al.}(2017)\citenamefont {Shao},
  \citenamefont {Qin}, \citenamefont {Capponi}, \citenamefont {Chesi},
  \citenamefont {Meng},\ and\ \citenamefont {Sandvik}}]{Nearly2017Shao}%
  \BibitemOpen
\bibfield  {journal} {  }\bibfield  {author} {\bibinfo {author} {\bibfnamefont
  {H.}~\bibnamefont {Shao}}, \bibinfo {author} {\bibfnamefont {Y.~Q.}\
  \bibnamefont {Qin}}, \bibinfo {author} {\bibfnamefont {S.}~\bibnamefont
  {Capponi}}, \bibinfo {author} {\bibfnamefont {S.}~\bibnamefont {Chesi}},
  \bibinfo {author} {\bibfnamefont {Z.~Y.}\ \bibnamefont {Meng}}, \ and\
  \bibinfo {author} {\bibfnamefont {A.~W.}\ \bibnamefont {Sandvik}},\ }\href
  {\doibase 10.1103/PhysRevX.7.041072} {\bibfield  {journal} {\bibinfo
  {journal} {Phys. Rev. X}\ }\textbf {\bibinfo {volume} {7}},\ \bibinfo {pages}
  {041072} (\bibinfo {year} {2017})}\BibitemShut {NoStop}%
\bibitem [{\citenamefont {Xu}\ \emph {et~al.}(2019)\citenamefont {Xu},
  \citenamefont {Xiong}, \citenamefont {Wu},\ and\ \citenamefont
  {Yao}}]{Spin2019yining}%
  \BibitemOpen
  \bibfield  {author} {\bibinfo {author} {\bibfnamefont {Y.}~\bibnamefont
  {Xu}}, \bibinfo {author} {\bibfnamefont {Z.}~\bibnamefont {Xiong}}, \bibinfo
  {author} {\bibfnamefont {H.-Q.}\ \bibnamefont {Wu}}, \ and\ \bibinfo {author}
  {\bibfnamefont {D.-X.}\ \bibnamefont {Yao}},\ }\href {\doibase
  10.1103/PhysRevB.99.085112} {\bibfield  {journal} {\bibinfo  {journal} {Phys.
  Rev. B}\ }\textbf {\bibinfo {volume} {99}},\ \bibinfo {pages} {085112}
  (\bibinfo {year} {2019})}\BibitemShut {NoStop}%
\bibitem [{\citenamefont {Sun}\ \emph {et~al.}(2018)\citenamefont {Sun},
  \citenamefont {Wang}, \citenamefont {Fang}, \citenamefont {Qi}, \citenamefont
  {Cheng},\ and\ \citenamefont {Meng}}]{Dynamical2018Sun}%
  \BibitemOpen
  \bibfield  {author} {\bibinfo {author} {\bibfnamefont {G.-Y.}\ \bibnamefont
  {Sun}}, \bibinfo {author} {\bibfnamefont {Y.-C.}\ \bibnamefont {Wang}},
  \bibinfo {author} {\bibfnamefont {C.}~\bibnamefont {Fang}}, \bibinfo {author}
  {\bibfnamefont {Y.}~\bibnamefont {Qi}}, \bibinfo {author} {\bibfnamefont
  {M.}~\bibnamefont {Cheng}}, \ and\ \bibinfo {author} {\bibfnamefont {Z.~Y.}\
  \bibnamefont {Meng}},\ }\href {\doibase 10.1103/PhysRevLett.121.077201}
  {\bibfield  {journal} {\bibinfo  {journal} {Phys. Rev. Lett.}\ }\textbf
  {\bibinfo {volume} {121}},\ \bibinfo {pages} {077201} (\bibinfo {year}
  {2018})}\BibitemShut {NoStop}%
\bibitem [{\citenamefont {Ma}\ \emph {et~al.}(2018)\citenamefont {Ma},
  \citenamefont {Sun}, \citenamefont {You}, \citenamefont {Xu}, \citenamefont
  {Vishwanath}, \citenamefont {Sandvik},\ and\ \citenamefont
  {Meng}}]{NSMA2018}%
  \BibitemOpen
  \bibfield  {author} {\bibinfo {author} {\bibfnamefont {N.}~\bibnamefont
  {Ma}}, \bibinfo {author} {\bibfnamefont {G.-Y.}\ \bibnamefont {Sun}},
  \bibinfo {author} {\bibfnamefont {Y.-Z.}\ \bibnamefont {You}}, \bibinfo
  {author} {\bibfnamefont {C.}~\bibnamefont {Xu}}, \bibinfo {author}
  {\bibfnamefont {A.}~\bibnamefont {Vishwanath}}, \bibinfo {author}
  {\bibfnamefont {A.~W.}\ \bibnamefont {Sandvik}}, \ and\ \bibinfo {author}
  {\bibfnamefont {Z.~Y.}\ \bibnamefont {Meng}},\ }\href {\doibase
  10.1103/PhysRevB.98.174421} {\bibfield  {journal} {\bibinfo  {journal} {Phys.
  Rev. B}\ }\textbf {\bibinfo {volume} {98}},\ \bibinfo {pages} {174421}
  (\bibinfo {year} {2018})}\BibitemShut {NoStop}%
\bibitem [{\citenamefont {Huang}\ \emph {et~al.}(2018)\citenamefont {Huang},
  \citenamefont {Deng}, \citenamefont {Wan},\ and\ \citenamefont
  {Meng}}]{CJHuang2018}%
  \BibitemOpen
  \bibfield  {author} {\bibinfo {author} {\bibfnamefont {C.-J.}\ \bibnamefont
  {Huang}}, \bibinfo {author} {\bibfnamefont {Y.}~\bibnamefont {Deng}},
  \bibinfo {author} {\bibfnamefont {Y.}~\bibnamefont {Wan}}, \ and\ \bibinfo
  {author} {\bibfnamefont {Z.~Y.}\ \bibnamefont {Meng}},\ }\href {\doibase
  10.1103/PhysRevLett.120.167202} {\bibfield  {journal} {\bibinfo  {journal}
  {Phys. Rev. Lett.}\ }\textbf {\bibinfo {volume} {120}},\ \bibinfo {pages}
  {167202} (\bibinfo {year} {2018})}\BibitemShut {NoStop}%
\bibitem [{\citenamefont {Ma}\ \emph {et~al.}(2019)\citenamefont {Ma},
  \citenamefont {You},\ and\ \citenamefont {Meng}}]{NVMa2019}%
  \BibitemOpen
  \bibfield  {author} {\bibinfo {author} {\bibfnamefont {N.}~\bibnamefont
  {Ma}}, \bibinfo {author} {\bibfnamefont {Y.-Z.}\ \bibnamefont {You}}, \ and\
  \bibinfo {author} {\bibfnamefont {Z.~Y.}\ \bibnamefont {Meng}},\ }\href
  {\doibase 10.1103/PhysRevLett.122.175701} {\bibfield  {journal} {\bibinfo
  {journal} {Phys. Rev. Lett.}\ }\textbf {\bibinfo {volume} {122}},\ \bibinfo
  {pages} {175701} (\bibinfo {year} {2019})}\BibitemShut {NoStop}%
\bibitem [{\citenamefont {{Yan}}\ \emph {et~al.}()\citenamefont {{Yan}},
  \citenamefont {{Wang}}, \citenamefont {{Ma}}, \citenamefont {{Qi}},\ and\
  \citenamefont {{Meng}}}]{ZhengYan2020}%
  \BibitemOpen
  \bibfield  {author} {\bibinfo {author} {\bibfnamefont {Z.}~\bibnamefont
  {{Yan}}}, \bibinfo {author} {\bibfnamefont {Y.-C.}\ \bibnamefont {{Wang}}},
  \bibinfo {author} {\bibfnamefont {N.}~\bibnamefont {{Ma}}}, \bibinfo {author}
  {\bibfnamefont {Y.}~\bibnamefont {{Qi}}}, \ and\ \bibinfo {author}
  {\bibfnamefont {Z.~Y.}\ \bibnamefont {{Meng}}},\ }\href {\doibase
  10.1038/s41535-021-00338-1} {\bibfield  {journal} {\bibinfo  {journal} {npj
  Quantum Mater.}\ }\textbf {\bibinfo {volume} {6}},\ \bibinfo {pages}
  {39}}\BibitemShut {NoStop}%
\bibitem [{\citenamefont {{Wang}}\ \emph {et~al.}(2020)\citenamefont {{Wang}},
  \citenamefont {{Cheng}}, \citenamefont {{Witczak-Krempa}},\ and\
  \citenamefont {{Meng}}}]{YCWang2020}%
  \BibitemOpen
  \bibfield  {author} {\bibinfo {author} {\bibfnamefont {Y.-C.}\ \bibnamefont
  {{Wang}}}, \bibinfo {author} {\bibfnamefont {M.}~\bibnamefont {{Cheng}}},
  \bibinfo {author} {\bibfnamefont {W.}~\bibnamefont {{Witczak-Krempa}}}, \
  and\ \bibinfo {author} {\bibfnamefont {Z.~Y.}\ \bibnamefont {{Meng}}},\
  }\href@noop {} {\bibfield  {journal} {\bibinfo  {journal} {arXiv e-prints}\
  ,\ \bibinfo {eid} {arXiv:2005.07337}} (\bibinfo {year} {2020})},\ \Eprint
  {http://arxiv.org/abs/2005.07337} {arXiv:2005.07337 [cond-mat.str-el]}
  \BibitemShut {NoStop}%
\bibitem [{\citenamefont {Hu}\ \emph {et~al.}(2020)\citenamefont {Hu},
  \citenamefont {Ma}, \citenamefont {Liao}, \citenamefont {Li}, \citenamefont
  {Ma}, \citenamefont {Cui}, \citenamefont {Shangguan}, \citenamefont {Huang},
  \citenamefont {Qi}, \citenamefont {Li}, \citenamefont {Meng}, \citenamefont
  {Wen},\ and\ \citenamefont {Yu}}]{BKT2020}%
  \BibitemOpen
  \bibfield  {author} {\bibinfo {author} {\bibfnamefont {Z.}~\bibnamefont
  {Hu}}, \bibinfo {author} {\bibfnamefont {Z.}~\bibnamefont {Ma}}, \bibinfo
  {author} {\bibfnamefont {Y.-D.}\ \bibnamefont {Liao}}, \bibinfo {author}
  {\bibfnamefont {H.}~\bibnamefont {Li}}, \bibinfo {author} {\bibfnamefont
  {C.}~\bibnamefont {Ma}}, \bibinfo {author} {\bibfnamefont {Y.}~\bibnamefont
  {Cui}}, \bibinfo {author} {\bibfnamefont {Y.}~\bibnamefont {Shangguan}},
  \bibinfo {author} {\bibfnamefont {Z.}~\bibnamefont {Huang}}, \bibinfo
  {author} {\bibfnamefont {Y.}~\bibnamefont {Qi}}, \bibinfo {author}
  {\bibfnamefont {W.}~\bibnamefont {Li}}, \bibinfo {author} {\bibfnamefont
  {Z.~Y.}\ \bibnamefont {Meng}}, \bibinfo {author} {\bibfnamefont
  {J.}~\bibnamefont {Wen}}, \ and\ \bibinfo {author} {\bibfnamefont
  {W.}~\bibnamefont {Yu}},\ }\href {\doibase 10.1038/s41467-020-19380-x}
  {\bibfield  {journal} {\bibinfo  {journal} {Nat. Commun.}\ }\textbf {\bibinfo
  {volume} {11}},\ \bibinfo {pages} {5631} (\bibinfo {year}
  {2020})}\BibitemShut {NoStop}%
\bibitem [{\citenamefont {Wang}\ \emph {et~al.}(2021)\citenamefont {Wang},
  \citenamefont {Yan}, \citenamefont {Wang}, \citenamefont {Qi},\ and\
  \citenamefont {Meng}}]{YCWang2021}%
  \BibitemOpen
  \bibfield  {author} {\bibinfo {author} {\bibfnamefont {Y.-C.}\ \bibnamefont
  {Wang}}, \bibinfo {author} {\bibfnamefont {Z.}~\bibnamefont {Yan}}, \bibinfo
  {author} {\bibfnamefont {C.}~\bibnamefont {Wang}}, \bibinfo {author}
  {\bibfnamefont {Y.}~\bibnamefont {Qi}}, \ and\ \bibinfo {author}
  {\bibfnamefont {Z.~Y.}\ \bibnamefont {Meng}},\ }\href {\doibase
  10.1103/PhysRevB.103.014408} {\bibfield  {journal} {\bibinfo  {journal}
  {Phys. Rev. B}\ }\textbf {\bibinfo {volume} {103}},\ \bibinfo {pages}
  {014408} (\bibinfo {year} {2021})}\BibitemShut {NoStop}%
\bibitem [{\citenamefont {Sandvik}(1999)}]{sandvik1999}%
  \BibitemOpen
  \bibfield  {author} {\bibinfo {author} {\bibfnamefont {A.~W.}\ \bibnamefont
  {Sandvik}},\ }\href {\doibase 10.1103/PhysRevLett.83.3069} {\bibfield
  {journal} {\bibinfo  {journal} {Phys. Rev. Lett.}\ }\textbf {\bibinfo
  {volume} {83}},\ \bibinfo {pages} {3069} (\bibinfo {year}
  {1999})}\BibitemShut {NoStop}%
\bibitem [{\citenamefont {Essler}\ and\ \citenamefont
  {Konik}(2005)}]{essler-konik}%
  \BibitemOpen
  \bibfield  {author} {\bibinfo {author} {\bibfnamefont {F.~H.~L.}\
  \bibnamefont {Essler}}\ and\ \bibinfo {author} {\bibfnamefont {R.~M.}\
  \bibnamefont {Konik}},\ }in\ \href@noop {} {\emph {\bibinfo {booktitle} {From
  Fields to Strings: Circumnavigating Theoretical Physics: Ian Kogan Memorial
  Collection (in 3 Vols)}}},\ Vol.~\bibinfo {volume} {1},\ \bibinfo {editor}
  {edited by\ \bibinfo {editor} {\bibfnamefont {M.}~\bibnamefont {Shifman}}}\
  (\bibinfo  {publisher} {World Scientific Publishing Co. Pte. Ltd},\ \bibinfo
  {year} {2005})\ pp.\ \bibinfo {pages} {684--830}\BibitemShut {NoStop}%
\bibitem [{\citenamefont {Hikihara}\ and\ \citenamefont
  {Starykh}(2010)}]{hikihara2010}%
  \BibitemOpen
  \bibfield  {author} {\bibinfo {author} {\bibfnamefont {T.}~\bibnamefont
  {Hikihara}}\ and\ \bibinfo {author} {\bibfnamefont {O.~A.}\ \bibnamefont
  {Starykh}},\ }\href {\doibase 10.1103/PhysRevB.81.064432} {\bibfield
  {journal} {\bibinfo  {journal} {Phys. Rev. B}\ }\textbf {\bibinfo {volume}
  {81}},\ \bibinfo {pages} {064432} (\bibinfo {year} {2010})}\BibitemShut
  {NoStop}%
\bibitem [{\citenamefont {Lukyanov}\ and\ \citenamefont
  {Zamolodchikov}(1997)}]{Lukyanov1997}%
  \BibitemOpen
  \bibfield  {author} {\bibinfo {author} {\bibfnamefont {S.}~\bibnamefont
  {Lukyanov}}\ and\ \bibinfo {author} {\bibfnamefont {A.}~\bibnamefont
  {Zamolodchikov}},\ }\href {\doibase
  https://doi.org/10.1016/S0550-3213(97)00123-5} {\bibfield  {journal}
  {\bibinfo  {journal} {Nucl. Phys. B}\ }\textbf {\bibinfo {volume} {493}},\
  \bibinfo {pages} {571 } (\bibinfo {year} {1997})}\BibitemShut {NoStop}%
\end{thebibliography}%
	\bibliographystyle{apsrev4-1}

	\setcounter{page}{1}
	\setcounter{equation}{0}
	\setcounter{figure}{0}
	\renewcommand{\theequation}{S\arabic{equation}}
	\renewcommand{\thefigure}{S\arabic{figure}}
	
	\newpage
	
\begin{widetext}
	\section{Supplemental Material}
	
	\centerline{\bf Amplitude Mode in Quantum Magnets via Dimensional Crossover}
	\vskip3mm
	
	\centerline{}
	
	\section{QMC-SAC Scheme}
	\label{app:SAC}
	The relationship between the imaginary-time correlation function of an operator $\hat{O}$, $G(\tau)=\langle\hat{O}(\tau)\hat{O}(0)\rangle$, and its corresponding spectral function, $A(\omega)$, can be given as
	
	\begin{equation}
		G(\tau)=\int_{-\infty}^{\infty} K(\tau,\omega)A(\omega)d\omega,
		\label{aeq:eq1}
	\end{equation}
	where the kernel, $K(\tau,\omega)$, depends on the type of the spectral function. For the bosonic case, there is a relation, $A(-\omega)=e^{-\beta\omega}A(\omega)$, between the spectral function at positive and negative frequency. Therefore, we restrict the integral in Eq.~\eqref{aeq:eq1} to the positive frequencies by applying the kernel
	\begin{equation}
		K(\tau)=\frac{1}{\pi}(e^{-\tau\omega}+e^{-(\beta-\tau)\omega}).
		\label{aeq:eq2}
	\end{equation}
	
	From the point of the SAC process, we use the normalization $G(0)=1$ to work with a spectral function that is itself normalized to unity on the positive frequency axis. Thus, Eq.\eqref{aeq:eq1} becomes
	\begin{equation}
		G(\tau)=\int_{0}^{\infty} \frac{e^{\tau\omega}+e^{-(\beta-\tau)\omega}}{1+e^{-\beta\omega}}B(\omega)d\omega,
		\label{aeq:eq3}
	\end{equation}
	where $B(\omega)=A(\omega)(1+e^{-\beta\omega})$ is the real-frequency spectral function. Hence, $\int_{0}^{\infty}d\omega B(\omega)=1$. 
	
	Practically, $B(\omega)$ is parameterized by a large number of equal-amplitude $\delta$ functions, which is sampled at location in a frequency continuum. The number of $\delta$ functions we used are $5000$ in both Fig.~\bl{2} and Fig.~\bl{3}. Therefore, $B(\omega)=\sum_{i=0}^{N_{\omega}-1}a_i\delta(\omega-\omega_i)$. Then, we update location of these $\delta$ functions in a Metropolis process using the likelihood function 
	\begin{equation}
		P(B)\propto \mathrm{exp}\left(-\frac{\mathrm{\chi}^2}{2\Theta}\right),
		\label{aeq:eq4}
	\end{equation}
	where $\Theta$ is the sampling temperature. And $\mathrm{\chi}^2$ denotes the goodness of fit, which describes the relation between correlation function obtained from QMC measurement, $G(\tau)$, and from Eq.~\eqref{aeq:eq3}, $G'(\tau)$. $\mathrm{\chi}^2$ is defined as
	\begin{equation}
		\mathrm{\chi}^2=\sum_{i,j}[G'(\tau_{i})-\overline{G}(\tau_{i})]C_{ij}^{-1}[G'(\tau_{j})-\overline{G}(\tau_{j})],
		\label{aeq:eq5}
	\end{equation}
	where $G'(\tau_{i})$ is obtained from the current spectral function by Eq.~\eqref{aeq:eq3} and $\overline{G}(\tau_{i})$ denotes the statistical average of QMC measurement. $C_{ij}$ is the covariance matrix element of the QMC data
	\begin{equation}
		C_{ij}=\frac{1}{N_b(N_b-1)}\sum^{N_b}_{b=1}[G^{b}(\tau_{i})-\overline{G}(\tau_{i})][G^{b}(\tau_{j})-\overline{G}(\tau_{j})].
		\label{aeq:eq6}
	\end{equation}
	Here $N_b$ refers to the number of bins in the QMC measurement, which is $128$ in this study. 
	
	Selecting $\Theta$ matters in the SAC process. We adopt the temperature-adjustment scheme devised in Ref.\cite{Nearly2017Shao}. In this scheme, a simulated annealing procedure is used to find the minimum value $\mathrm{\chi}_{min}^2$. After this initial step, $\Theta$ is adjusted to make the average $\mathrm{\chi}^2$ in the final sampling process for the spectral function satisfy 
	\begin{equation}
		\langle\mathrm{\chi}^2\rangle\approx\mathrm{\chi}_{min}^2+\sqrt{2N_{\tau}}.
		\label{aeq:eq7}
	\end{equation}
	Here $N_{\tau}$ is the size of time point set used in the SAC process, which we will discuses below. With a suitable $\Theta$, a smooth averaged spectral function can be obtained and the final spectral function is the ensemble average of the Metropolis process within the confrontational space of $\{a_i, \omega_i\}$, as explained in Refs.~\cite{Constrained2016Sandvik,Amplitude2017Qin}.
	
	The spectral function is `good' when $\langle\mathrm{\chi}^2\rangle$ closes to one. To do so, it is useful to find a suitable time point set $\{G(\tau_i)\}$ that input into the SAC process. Here, those $G(\tau)$ with relative statistic errors larger than $10^{-1}$ should not be considered in the SAC process since the ill-posed nature of the inverse Laplace transform. Besides, for a dataset of $G(\tau)$ from the QMC measurement, each time point is chosen and put into the SAC process with the probability
	\begin{equation}
		\begin{aligned}
			P(\tau)=\left\{
			\begin{aligned}
				& 1, \quad \tau=\frac{A_{n}^{2}d\tau}{4}\\
				& p, \quad \tau\neq\frac{A_{n}^{2}d\tau}{4}
			\end{aligned}
			\right.
		\end{aligned}
		\label{aeq:eq8}
	\end{equation}
	Here, $d\tau$ is the resolution of $G(\tau)$ in the QMC measurement. And $A_{n}=\alpha n$ is a sequence of $n=0,1,2...$ and parameter $\alpha$. $0<p<1$ is a constant. By changing parameters $\alpha$ and $p$, one can construct a suitable $\{G(\tau_i)\}$ and make $\langle\mathrm{\chi}^2\rangle=1\pm0.1$. Finally, we apply $N_{\tau}\approx50$ in the SAC process. 
	
	\section{Fitting the excitation gap}
	\label{app:GAP}
	In the Fig.~\bl{4} of the main text, we performed finite size scaling of the four different excitation gaps. These gap values are obtained from fitting their normalized imaginary-time correlation function $G(\tau)$ with $G_{\text{fit}}(\tau)=be^{-\omega\tau}$, in which $\omega$ refers to the low-energy mode frequency and $b$ is the fitting parameter.
	
	In order to achieve a good fitting, we first find a large $\tau_{\text{max}}$ with a small relative error ($<0.25$) in $G(\tau)$ obtained from QMC. For example, in Fig.\ref{fig:figS1}(a), we prefer $\tau_{\text{max}}=23.04$, at which the relative error of $G(\tau)$ is $0.224$. Secondly, we choice a $\tau_{\text{min}}$ in $G(\tau)$, which would give a good fitting between $\tau_{\text{min}}$ and $\tau_{\text{max}}$. We use $\tau_{\text{min}}=5.06$ and obtain $G_{\text{fit}}(\tau)=0.575e^{-0.306\tau}$, which has been plotted as the blue line in Fig.\ref{fig:figS1}(a). Then, we shift $\tau_{\text{min}}$ to four closeby imaginary-time points and obtain four other values of $\omega$. For $G_{S^{x}}(\tau)$ in Fig.\ref{fig:figS1}(a), these four points are $\tau_1=4.62$, $\tau_2=4.84$, $\tau_3=5.29$ and $\tau_4=5.52$. Finally, we apply the average over these five different $\omega$ as the mode frequency, and their sample standard deviation as the error bar in Fig.~\bl{4}, which is $0.306\pm0.028$ for $A_{S^{x}}$ with $g=0.1$, $h=0$ and system size $L=36$ at $\textbf{k}=(\pi,\pi/2)$. With this method, we have also plotted the normalized imaginary-time correlation functions of spin (or bond) operator and our fitting results in Fig.\ref{fig:figS1}. 
	
	\begin{figure*}[htp]
		\centering
		\includegraphics[width=\textwidth]{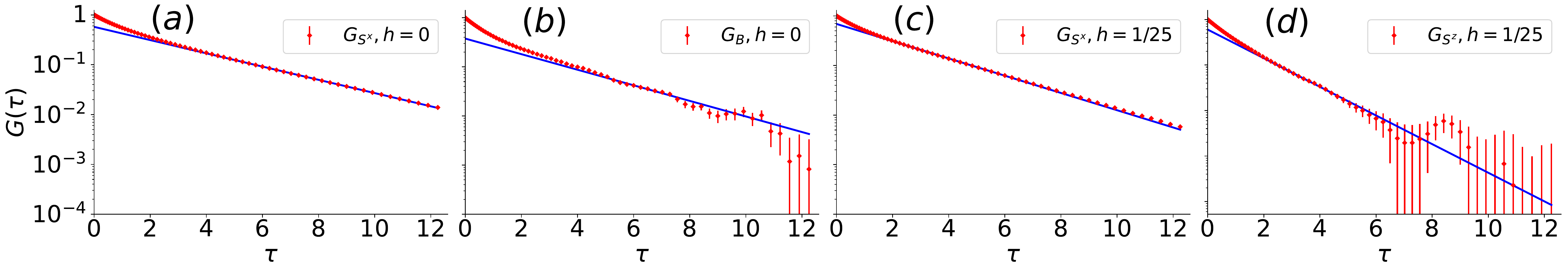}
		\caption{Scheme of the imaginary-time correlation fitting. (a-d) show the normalized imaginary-time correlation of corresponding operator respectively [$G_{S^{z}}(\tau)$, $G_{B}(\tau)$, $G_{S^{z}}(\tau)$ and $G_{S^{x}}(\tau)$], with $g=0.1$ and system size $L=36$ at $\textbf{k}=(\pi,\pi/2)$. The red dots are QMC data and the blue lines are our fitting curves.}	
		\label{fig:figS1}
	\end{figure*}
	
	\section{bond-bond correlation measurement}
	\label{app:DOT}
	To measure the bond-bond correlation $G_{B}$ (the scalar mode), we design the follow correlation functions, by using $O_{3}$ spin-rotational symmetry, 
	\begin{equation}
		\begin{aligned}
			G_{B}(i-j,\tau)&=\langle B_{i}(\tau)B_{j}(0)\rangle\\
			&=\langle[\mathbf{S}_{i+\hat{x}}(\tau)\cdot\mathbf{S}_{i}(\tau)][\mathbf{S}_{j+\hat{x}(0)}\cdot \mathbf{S}_{j}(0)]\rangle\\
			&=3\langle[S^{\mathrm{z}}_{i+\hat{x}}(\tau)S^{\mathrm{z}}_{i}(\tau)][\mathbf{S}_{j+\hat{x}(0)}\cdot\mathbf{S}_{j}(0)]\rangle\\
			&=3\langle S^{\mathrm{z}}_{i+\hat{x}}(\tau)S^{\mathrm{z}}_{i}(\tau)S^{\mathrm{z}}_{j+\hat{x}}(0) S^{\mathrm{z}}_{j}(0)\rangle
			+\frac{3}{2}\langle [S^{\mathrm{z}}_{i+\hat{x}}(\tau)S^{\mathrm{z}}_{i}(\tau)][S^{+}_{j+\hat{x}}(0)S^{-}_{j}(0)+S^{-}_{j+\hat{x}}(0) S^{+}_{j}(0)]\rangle\\
			&=3\langle B^{\mathrm{z}}_{i}(\tau)B^{\mathrm{z}}_{j}(0) \rangle+3\langle B^{\mathrm{z}}_{i}(\tau)B^{\pm}_{j}(0) \rangle\\
			&=3G_{B^{\mathrm{z}}}(i-j,\tau)+3G_{B^{\mathrm{c}}}(i-j,\tau).
		\end{aligned}
		\label{aeq:eq9}
	\end{equation} 
	Here, $B^{\mathrm{z}}_{i}=S^{\mathrm{z}}_{i+\hat{x}}S^{\mathrm{z}}_{i}$ and $B^{\mathrm{\pm}}_{i}=\frac{1}{2}(S^{+}_{i+\hat{x}} S^{-}_{i}+S^{-}_{i+\hat{x}} S^{+}_{i})$ are the longitudinal and transverse  components of the bond-bond correlation. As shown in Eq.~\eqref{aeq:eq9}, $G_{B}$ contains two different parts, which are the diagonal term $G_{B^{\mathrm{z}}}(i-j,\tau)=\langle B^{\mathrm{z}}_{i}(\tau)B^{\mathrm{z}}_{j}(0) \rangle$ and the cross term $G_{B^{\mathrm{c}}}(i-j,\tau)=\langle B^{\mathrm{z}}_{i}(\tau)B^{\pm}_{j}(0) \rangle$. 
	When applying the $\{S^{\mathrm{z}}\}$ basis in SSE-QMC simulation~\cite{Computational2010Sandvik,Quantum2002Sandvik,Generalized2005Alet}, the measurement of the $G_{B^{\mathrm{z}}}$ term is simple since it is also in the eigenbasis of the $B^{\mathrm{z}}_{i}$ operator. For a given static configuration space, one can calculate the value of $B^{\mathrm{z}}_{i}$ directly. However, the measurement of the cross term $G_{B^{\mathrm{c}}}$ is difficult, since the $\{S^{\mathrm{z}}\}$ basis is not the eigenbasis of $B^{\mathrm{\pm}}_{i}$ and the cross term measurement cannot be directly applied in a given static configuration space. In fact, in the previous QMC works about the measurement of bond-bond correlations, it is usually the $G_{B^{\mathrm{z}}}$ term that has been measured~\cite{Amplitude2017Qin,Lohoefer2017}, for the sake of simplicity as mentioned above.
	
	However, in our case of the coupled spin chains, we have seen that the measurement of $G_{B^{\mathrm{z}}}$ will mix with the phase mode of spin wave and generate additional spurious features close to the $\Gamma$ point (although the breather mode between $M \to X_1$ close to the 1D limit is kept intact), and it is only in the full measurement of $G_{B^{\mathrm{z}}}+G_{B^{\mathrm{c}}}$, including the contribution of the cross term, that the true scalar mode spectra are revealed. Below we explain how such full measurement of $G_{B}$ is implemented in the QMC simulation.
	
	To measure the cross term, it is convenience to bring our attention back to the updates of configuration spaces in the SSE-QMC simulation. There are two important updates, the diagonal update and the directed loop update. The former is about inserting and removing the diagonal operators in the configuration spaces while the latter can be viewed as the creation, annihilation, and movement of the off-diagonal operator in the configuration space. For example, in our model, the Hamiltonian (Eq.~(\bl{1})) only includes two kinds of off-diagonal term, $S_{i}^{+}S_{j}^{-}$ and $S_{i}^{-}S_{j}^{+}$. Hence, the direct loop update process can be understood as follows. A pair of off-diagonal operators, $S_{i}^{+}$ and $S_{i}^{-}$, are created on a random site $i$ at a random time $\tau$ in the configurations space. One of the operators is fixed while the other moves in the configuration space according to the detail balance condition. When these two operators meet each other again, they annihilate and the loop is construed. Therefore, within this picture, to measure the correlation about $S_{i}^{+}$ and $S_{i}^{-}$ is about tracing the movement of off-diagonal operators in the configuration space when constructing loops in the directed loop update process.
	
	\begin{figure*}[htp]
		\centering
		\includegraphics[width=\textwidth]{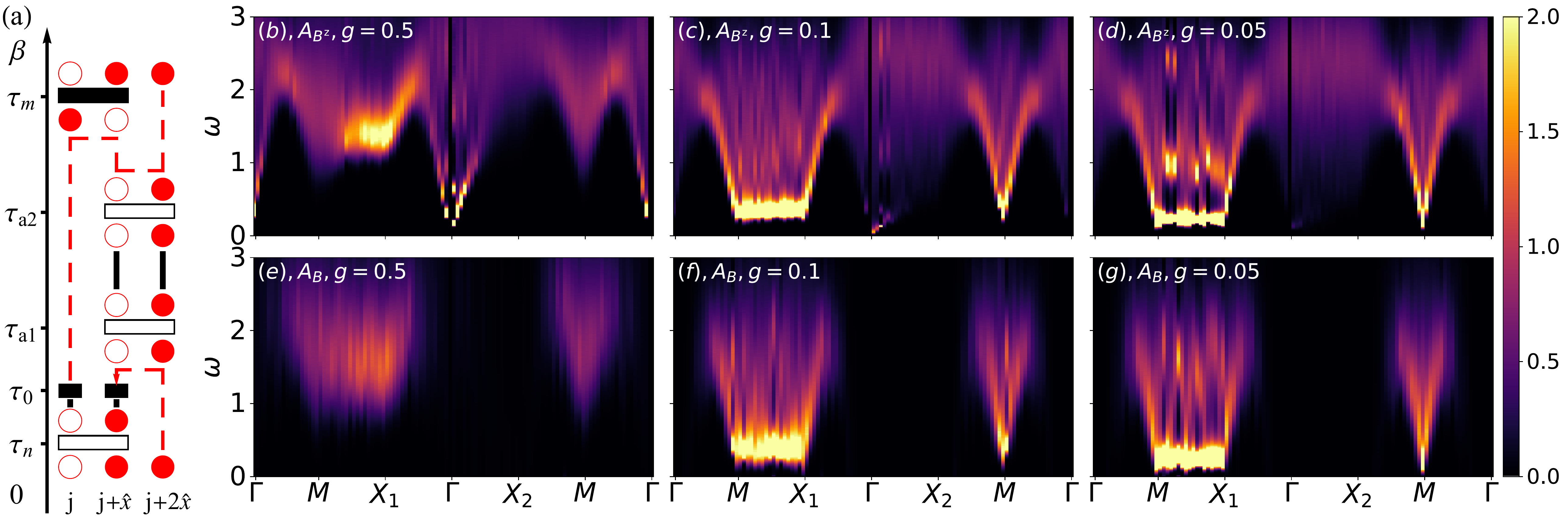}
		\caption{(a) The schematic diagram of the cross term $G_{B^{\mathrm{c}}}$ measurement in the configuration space of SSE-QMC simulation. The red solid (hollow) circle refers to the spin up (down). And the big solid (hollow) rectangles is the (off-)diagonal operator, such as $S^{\mathrm{z}}_{i}S^{\mathrm{z}}_{j}$ for diagonal and $S^{\mathrm{+}}_{i}S^{\mathrm{-}}_{j}$ for off-diagonal. The two small rectangles is the pair of operator created in the directed loop update process and the red dashed line is the path of the moving operator. Also, we plot the imaginary time axis pointing from $0$ to $\beta$, and map each operator to a space-time coordinate (like $(i,\tau_0)$). (b), (c), (d) and (e), (f), (g) show the spectra function of bond operators respectively, $A_{B^{\mathrm{z}}}(\mathbf{q},\omega)$ and $A_{B}(\mathbf{q},\omega)$ at $h=0$, at different values of $g=J/J_\perp$, with the system size $L=36$ and inverse temperature $\beta=4L$.}
		\label{fig:figS2}
	\end{figure*}
	
	In practice, starting from a given configuration space with $N_{h}$ operators, we first map $N_{h}$ operators to $N_{h}$ sorted random imaginary time point between $0$ and $\beta$. For example, as in Fig.~\ref{fig:figS2} (a), there are four bond operators. Hence, we map them to four imaginary time point, which are $\tau_{n}$, $\tau_{a1}$, $\tau_{a2}$, and $\tau_{m}$. Noted that each imaginary time point is selected randomly but fulfill that $0\leqslant\tau_{n}<\tau_{a1}<\tau_{a2}<\tau_{m}\leqslant\beta$. Therefore, one can deduce the spin state in the configuration space for a given $\tau$. When measuring the pure term $G_{B^{\mathrm{z}}}$, we select a imaginary time point $\tau_{0}$ randomly in the configuration space. And then, we deduce the corresponding spin state at $\tau_{0}$ and $\tau_{0}+\tau$ in the configuration space, and calculate the value of $\langle B^{\mathrm{z}}_{i}(\tau)B^{\mathrm{z}}_{j}(0) \rangle$. With a large number of sampling ($\approx10^4$ Monte Carlo step), the final result of $G_{B^{\mathrm{z}}}$ is given as the average. 
	
	For the cross term measurement, within the directed loop update in SSE-QMC simulation, we first select a vertex leg in the configuration space randomly. Then, we create a pair of off-diagonal operators at the random point $(j,\tau_{0})$ on this vertex leg. For example, in Fig.\ref{fig:figS2}(a), $S_{j}^{+}$ and $S_{j}^{-}$ are created at $(j,\tau_{0})$, which is on the vertex leg connecting the operators at $\tau_{n}$ and $\tau_{m}$. Since $\tau_{0}$ here is selected randomly in the region $[\tau_{n},\tau_{m}]$ instead of $[0,\beta]$, the creation here causes a weight $w^{l}_{mn}=\mathrm{mod}[(\tau_{m}-\tau_{n}),\beta]$ in our measurement, where $l$ is the index of constructing loop and $\mathrm{mod}$ calculation comes from the periodic boundary condition in the imaginary time.
	
	Then, we fix one of the inserted operators and move the other according to the detail balance condition. Meanwhile, we trace the path of the moving operator and monitor whether it arrives the point $(j+\hat{x},\tau_{0})$. If the moving operator arrives the point $(j+\hat{x},\tau_{0})$ (as in Fig.\ref{fig:figS2}(a) with the periodic boundary condition in time), we pause the construction of the loop and measure the $B^{\mathrm{z}}_{i}(\tau)$ relative to $(j,\tau_{0})$ in the current configuration space, in which the measurement result are recorded as $\mathrm{g}^{l}(i-j,\tau)$ and $l$ is the loop index. If the moving operator never meets $(j+\hat{x},\tau_{0})$, it means that $S^{+}_{j}S^{-}_{j+\hat{x}}$ or $S^{-}_{j}S^{+}_{j+\hat{x}}$ does not appear in the current loop construction. Therefore, the measurement result $\mathrm{g}^{l}(i-j,\tau)=0$. If the moving operator meet the fixed one, they annihilate and directed loop update completes. With large enough samplings ($\approx 10^{4}L$ loops in our simulation), the final measurement result of cross term can be read as
	\begin{equation}
		\begin{aligned}
			G_{B^{\mathrm{c}}}(i-j,\tau)=\frac{\sum_{l} w^{l}_{mn}\mathrm{g}^{l}(i-j,\tau)}{\sum_{l} w^{l}_{mn}}.
		\end{aligned}
		\label{aeq:eq10}
	\end{equation} 
	where the weights $w^{l}_{mn}$ is given above. In this way, both $G_{B^{\mathrm{z}}}$ and $G_{B^{\mathrm{c}}}$ are measured in the SSE-QMC and we obtain the full bond-bond correlation function as in Eq.~\eqref{aeq:eq9}.
	
	\begin{figure*}[htp]
		\centering
		\includegraphics[width=0.9\textwidth]{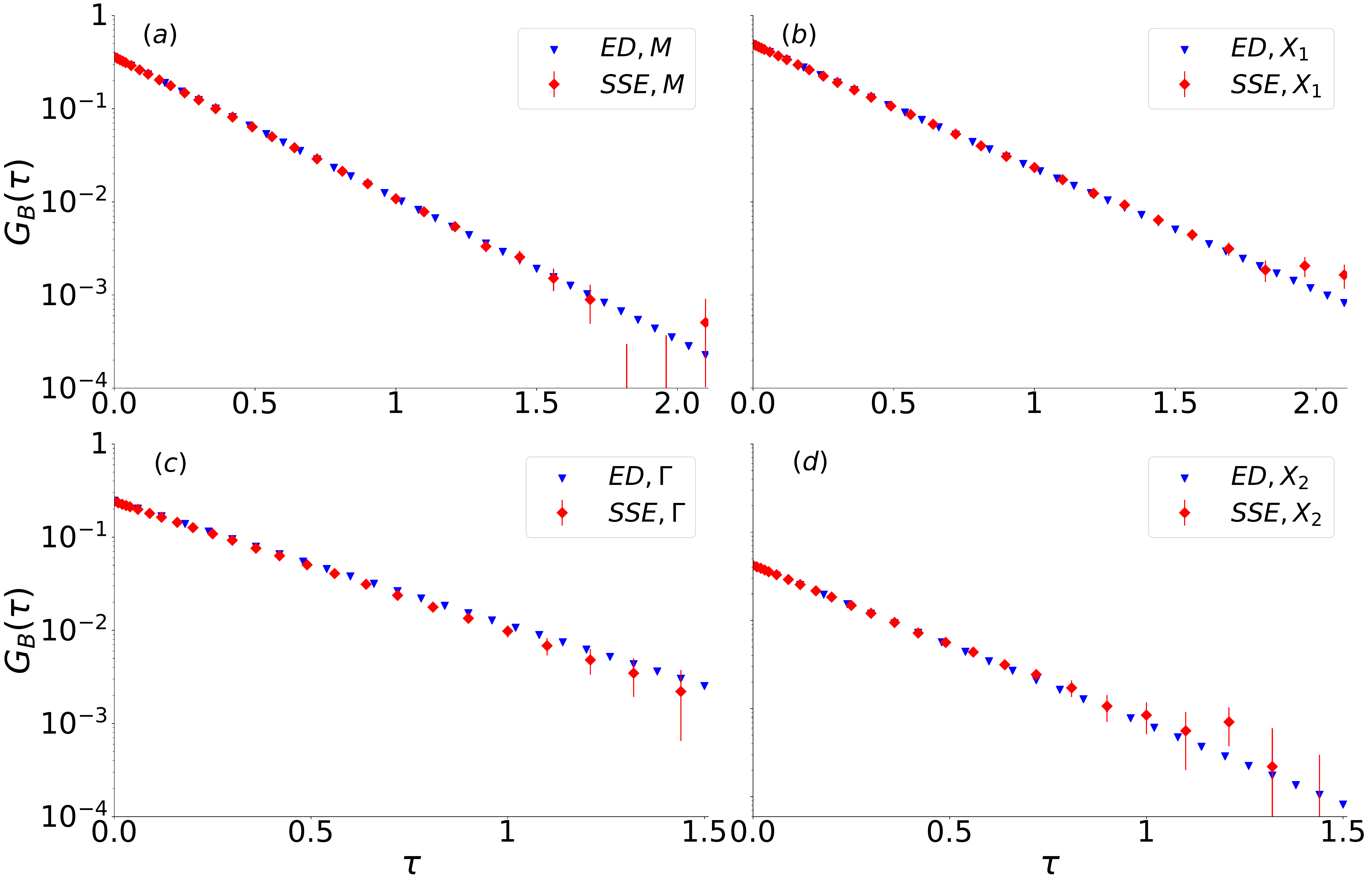}
		\caption{Comparison of the dynamic bond correlation from from exact diagonalization and QMC simulation as a function of imaginary time, the system parameters are $L=4$, $h=0$ and $g=1$ for $G_{B}(\tau)$ at (a) $M$, (b) $X_{1}$, (c) $\Gamma$ and (d) $X_{2}$ momenta.}
		\label{fig:figS3}
	\end{figure*}
	
	To make sure the measurement of the dynamic bond correlation is correct, we first compare the imaginary time results with the same measurements from the exact diagonalization (ED), and the comparison is shown in Fig.~\ref{fig:figS3}. This is a small system with $L=4$ and we plot the imaginary time decay of the full dynamic bond correlation at four momenta ($G_B(\Gamma,\tau)$, $G_B(X_1,\tau)$, $G_B(X_2,\tau)$ and $G_B(M,\tau)$), the ED imaginary time results are obtained from the Fourier transformation of Matsubara correlation function obtained by continued fraction expansion using Lanczos method. The match is perfect within Monte Carlos statistical errors.
	
	Meanwhile, in Fig.~\ref{fig:figS2} (b) - (g), we show the comparison of the spectral functions of bond-bond correlation $A_{B}$ ($A_{B^{\mathrm{z}}}$) in $B$ ($B^{\mathrm{z}}$) channels, respectively. It is clear that the main difference between $A_{B}$ and $A_{B^{z}}$ comes from the region close to $\Gamma$ points in Fig.\ref{fig:figS2}, especially at the 2D limit with $g=0.5$. In $A_{B^{\mathrm{z}}}$ (Fig.~\ref{fig:figS2} (b) and (c)), it shows a gapless mode at $\Gamma$ points, this is due to the mixing between the $B^{\mathrm{z}}$ and the phase mode of $S^{\mathrm{z}}$. And such artificial features disappear only the cross term $B^{c}$ is properly taken into the measurement and does not exist in $A_{B}$ (Fig.~\ref{fig:figS2} (e) and (f)). Although the first breather at the 1D limit between $M \to X_1$ are similar in Fig.~\ref{fig:figS2} (d) and (g) with $g=0.05$, in the main text, we use the full bond-bond correlation as the correct measurement.
	
	\section{Field theory treatment}
	In this section, we discuss the theoretic treatment, where the interchain coupling is treated within a mean-field and random-phase approximation while the intrachain (1D) physics are treated via bosonization and exact methods. For more details about this theory approach, see the review by Essler and Konik~\cite{essler-konik} and reference therein. In comparison with Ref.~\onlinecite{essler-konik}, here it is needed to generalize the theory analysis a bit in order to incorporate the external staggered field utilized in the simulations.
	
	Here, we will set the temperature $T=0$ and the system size $L=\infty$. Without loss of generality, we set the AFM order to be along the $S^z$ direction with an order parameter $m_0=(-1)^i \langle S_i^z\rangle$.
	
	\subsection{Self-consistent Mean-field Treatment}
	Here, we start from taking the self-consistent mean-field treatment for the interchain couplings ($J_{\perp}\sum_{\langle i,j \rangle_{\mathrm{y}}}\mathbf{S}_{i}\cdot\mathbf{S}_{j} $), while quantum fluctuations around the mean-field ground state will be considered later via a random-phase approximation (RPA). Within the mean-field treatment, $\langle S_i^z\rangle=m_0 (-1)^i$ and thus this interchain coupling turns into an effective staggered field for each spin chain ($h_{\rm eff}$) with a mean-field Hamiltonian
	\begin{equation}
		H_{MF} =J\sum_{\langle i,j \rangle_{\mathrm{x}}}\mathbf{S}_{i}\cdot \mathbf{S}_{j}-(h+h_{\rm eff})\sum_{i} (-1)^{i}S^{z}_{i},
	\end{equation} 
	where  
	\begin{equation}
		h_{\rm eff}= 2 J_{\perp} m_0.
		\label{eq:heff:definition}
	\end{equation} 
	and $h$ is the external staggered field.
	
	For this mean-field Hamiltonian, the 1D spin chains decouple from each other, and thus it can be solved via 1D exact/non-perturbative methods. At low-energy, the mean-field Hamiltonian of each spin chain can be reduced to the quantum sine-Gordon (SG) model  via bosonization with an Hamiltonian
	\ifdefined\details  \textcolor{red}{[Eq. (3.119) E-K] } \fi
	\begin{equation}
		H_{SG} =\int dx \left \{\frac{v}{16\pi} \left[\left(\partial_x\Phi \right)^2+\left(\partial_x\theta\right)^2\right]-\mu \cos\left(\beta \Phi\right)\right\}
	\end{equation}
	where $\beta=1/2$, $v=\pi J a_0/2$, and $\mu=c (h+h_{\rm eff}) a_0^{-1/2}$. The  constant $c$ here is $1/2$ and $a_0$ is the lattice constant, which will be set to unity ($a_0=1$). 
	In this bosonized form, the spin $S_i^z$ operator becomes
	\ifdefined\details  \textcolor{red}{[Eq. (3.122) E-K] } \fi
	\begin{equation}
		S^z_i\sim \frac{\partial_x \Phi }{4\pi}+ c (-1)^i \cos \left(\Phi/2\right),
	\end{equation}
	and the expectation value of the order parameter is $m_0=(-1)^i \langle S_i^z\rangle=c \langle\cos \left(\Phi/2\right)\rangle$.
	This expectation value was evaluated in Ref.~\onlinecite{Lukyanov1997} and is given by
	\ifdefined\details  \textcolor{red}{[Eq. (15) L-Z] } \fi
	\begin{align}
		m_0=&\frac{(2\pi J /v)^{1/3}}{6\sqrt{3}}
		\left[\frac{\Gamma(\frac{3}{4})}{\Gamma(\frac{1}{4})} \right]^{4/3}
		\left[\frac{\Gamma(\frac{1}{6})}{\Gamma(\frac{2}{3})} \right]^2
		\left(
		\frac{h + h_{\rm eff}}{J}
		\right)^{1/3}
		\label{eq:m0:SGM:LZ}
	\end{align}
	\ifdefined\details  \textcolor{red}
	{Compare with Ref.~\onlinecite{Lukyanov1997}: Notice that as shown in Eq. (5) of Ref.~\onlinecite{Lukyanov1997}, the velocity is set to unity. More importantly, the coefficient of $\cos$ term is $2\mu$ instead of $\mu$, and thus we need to replace $\mu$ in Ref.~\onlinecite{Lukyanov1997} by $\frac{\mu}{2}$ to follow our definition for the cosine term. From Eq.~(12), we get
		\begin{align}
			\Delta=&\frac{2}{\sqrt{\pi}}\frac{\Gamma(\xi/2)}{\Gamma(\frac{1+\xi}{2})}\left(\pi \frac{\Gamma(1-\beta^2)}{\Gamma(\beta^2)} \frac{\mu}{2}\right)^{1/(2-2\beta^2)}
			=\frac{2}{\sqrt{\pi}}\frac{\Gamma(\xi/2)}{\Gamma(\frac{1+\xi}{2})}\left(\frac{\pi}{2} \frac{\Gamma(\frac{1}{1+\xi})}{\Gamma(\frac{\xi}{1+\xi})} \mu\right)^{(1+\xi)/2}
		\end{align}
		Here, let us add the velocity back, whose value was set to unity in Ref.~\onlinecite{Lukyanov1997}. Because $v a_0$ sets the characteristic energy scale in the SG theory, 
		as we add $v$ and $a_0$ back, $\Delta$ on the l.h.s should be replaced with the dimensionless gap  $\Delta \to \frac{\Delta}{v/a_0}$, while the $\mu$ on the r.h.s. should be replaced by the dimensionless magnetic field strength $\mu \to \frac{c h}{v/a_0}$
		\begin{align}
			\frac{\Delta}{v/a_0}=\frac{2}{\sqrt{\pi}}\frac{\Gamma(\xi/2)}{\Gamma(\frac{1+\xi}{2})}\left(\frac{\pi}{2} \frac{\Gamma(\frac{1}{1+\xi})}{\Gamma(\frac{\xi}{1+\xi})} \frac{c h}{v/a_0}\right)^{(1+\xi)/2}
		\end{align}
		It is easy to check that if we set $v=a_0=1$, this dimensionless equation recovers the results of Ref.~\onlinecite{Lukyanov1997}.
		If we define dimensionless velocity $\tilde{v}= v/(J a_0)$, the equation becomes
		\begin{align}
			\frac{\Delta}{J}=\frac{2 \tilde{v}}{\sqrt{\pi}}\frac{\Gamma(\xi/2)}{\Gamma(\frac{1+\xi}{2})}\left(\frac{\pi c}{2 \tilde{v}} \frac{\Gamma(\frac{1}{1+\xi})}{\Gamma(\frac{\xi}{1+\xi})} \frac{h}{J}\right)^{(1+\xi)/2}
			\label{app:eq:dimensionlessgap}
		\end{align}
		which recovers Eq. 3.55 in Ref.~\onlinecite{essler-konik}.}
	
	\textcolor{red}{Now, let us look at the expectation value of the order parameter. From Eq.~(15) of Ref.~\onlinecite{Lukyanov1997},
		\begin{align}
			\langle e^{i \beta \Phi}\rangle
			=\frac{(1+\xi)\pi}{16\sin(\pi \xi)}\frac{\Gamma(1-\beta^2)}{\Gamma(\beta^2)}
			\left( 
			\frac{\Gamma(\frac{1+\xi}{2})\Gamma(1-\frac{\xi}{2})}{4\sqrt{\pi}}
			\right)^{2\beta^2-2} m^{2\beta^2}
		\end{align}
		At $\beta=1/2$ and $\chi=1/3$, we have
		\begin{align}
			\langle e^{i \beta \Phi}\rangle
			=\frac{\pi}{6\sqrt{3}}\frac{\Gamma(3/4)}{\Gamma(1/4)}
			\left( 
			\frac{\Gamma(2/3)\Gamma(5/6)}{4\sqrt{\pi}}
			\right)^{-3/2} \Delta^{1/2}
			=\frac{\pi}{6\sqrt{3}}\frac{\Gamma(3/4)}{\Gamma(1/4)}
			\left( 
			\frac{2}{\sqrt{\pi}}
			\frac{\Gamma(1/6)}{\Gamma(2/3)}
			\right)^{3/2} \Delta^{1/2}
		\end{align}
		In the last step, we used the fact that $\Gamma(1/6)\times\Gamma(5/6)=2\pi$. Now, we add back $v$ and $a_0$.
		\begin{align}
			m_0=\frac{\pi}{6\sqrt{3}}\frac{\Gamma(3/4)}{\Gamma(1/4)}
			\left( 
			\frac{2}{\sqrt{\pi}}
			\frac{\Gamma(1/6)}{\Gamma(2/3)}
			\right)^{3/2} \left(\frac{\Delta}{v/a_0}\right)^{1/2}
		\end{align}
		Here, the l.h.s. is already a dimensionless quantity (the order parameter $m_0$), and thus there is no need to rescale it to make it dimensionless. For the r.h.s., we replace $\Delta$ with the dimensionless gap $\Delta/(v/a_0)$. At $v=1$ and $a_0=1$, the formula recovers the dimensionless form above. Using Eq.~\eqref{app:eq:dimensionlessgap}, we get
		\begin{align}
			m_0=&c  \langle \cos (\beta \Phi)\rangle
			= c \frac{\pi}{6\sqrt{3}}\frac{\Gamma(3/4)}{\Gamma(1/4)}
			\left( 
			\frac{2}{\sqrt{\pi}}
			\frac{\Gamma(1/6)}{\Gamma(2/3)}
			\right)^{3/2} 
			\left(
			\frac{2}{\sqrt{\pi}}\frac{\Gamma(1/6)}{\Gamma(2/3)}\left(\frac{\pi c}{2 \tilde{v}} \frac{\Gamma(3/4)}{\Gamma(1/4)} \frac{h}{J}\right)^{2/3}
			\right)^{1/2}
			\nonumber\\
			=&\frac{1}{2}\frac{(2\pi/\tilde{v})^{1/3}}{3\sqrt{3}}
			\left[\frac{\Gamma(\frac{3}{4})}{\Gamma(\frac{1}{4})} \right]^{4/3}
			\left[\frac{\Gamma(\frac{1}{6})}{\Gamma(\frac{2}{3})} \right]^2
			\left(
			\frac{h}{J}
			\right)^{1/3}
		\end{align}
		Because $h_{eff}/J=2 g m_0$ and $\tilde{v}=\pi/2$, we get
		\begin{align}
			\frac{h_{eff}}{J}
			=&\frac{2^{2/3}}{3\sqrt{3}}
			\left[\frac{\Gamma(\frac{3}{4})}{\Gamma(\frac{1}{4})} \right]^{4/3}
			\left[\frac{\Gamma(\frac{1}{6})}{\Gamma(\frac{2}{3})} \right]^2
			g
			\left(
			\frac{h+h_{eff}}{J}
			\right)^{1/3}
		\end{align}
	}
	\fi
	
	Because $h_{\rm eff}= 2 J_{\perp} m_0$, Eqs.~\eqref{eq:heff:definition} and~\eqref{eq:m0:SGM:LZ} enforce a self-consistency condition for $h_{\rm eff}$
	\begin{align}
		\frac{h_{\rm eff}}{J}=\gamma_1 \frac{J_{\perp}}{J} \left(\frac{h+h_{\rm eff}}{J}\right)^{1/3},
		\label{eq:self:consistency:LZ}
	\end{align}
	where the coefficient 
	\begin{align}
		\gamma_1=\frac{2^{2/3}}{3\sqrt{3}}
		\left[\frac{\Gamma(\frac{3}{4})}{\Gamma(\frac{1}{4})} \right]^{4/3}
		\left[\frac{\Gamma(\frac{1}{6})}{\Gamma(\frac{2}{3})} \right]^2 
		\approx 1.215340
		\label{eq:SI:gamma:1}
	\end{align}
	Here, we used the fact that $v/J=\pi/2$.
	By solving cubic equation \eqref{eq:self:consistency:LZ}, the value of the effective staggered field ($h_{\rm eff}$) can be determined
	\begin{align}
		h_{\rm eff}=\frac{3 h}{x}\cos\left(\frac{\arccos x}{3}\right)
		\label{eq:heff:LZ/:part1}
	\end{align}
	where
	\begin{align}
		x=\frac{3\sqrt{3}}{2}\frac{h/J}{ \left( \gamma_1 \frac{J_\perp}{J}\right)^{3/2}} =\frac{3\sqrt{3}}{2}\frac{h/J}{ \left( \gamma_1 g\right)^{3/2}} 
		\label{eq:heff:LZ/:part2}
	\end{align}
	Notice that Eq.~\eqref{eq:heff:LZ/:part1} is defined for $x\le 1$ (weak field) due to the utilization of $\arccos x$. To extend it to $x>1$ (strong field), one can just replace 
	$\cos(\arccos(x)/3)$ with its
	analytic continuation $\cosh(\textrm{arccosh}(x)/3)$ .
	In the absence of external field $h=0$, the equation above reduces to $h_{\rm eff}= \left( \gamma_1 g\right)^{3/2}J$.
	
	Given relatively short chains used in our numerical study, $L = 20 - 36$, the derivation above does not account for the logarithmic corrections to the mass gap and other quantities. These corrections arise due to the marginally irrelevant interaction of chiral spin currents \cite{ao1999,essler-konik} and become important in the long chain limit \cite{sandvik1999}.
	
	\subsection{1D Spectral Functions}
	In this section, we compute the spectrum function of spin-spin and bond-bond correlators for the 1D mean-field Hamiltonian, ignoring any inter-chain quantum fluctuations, which will be studied in the next section. In the bosonized form, the SG model described above (with $\beta=1/2$) contains four gapped excitations. In notations of \cite{essler-konik} here 
	$\xi = 1/3$. Three of them (soliton, anti-soliton and the first breather $B_1$) shares the same energy gap $\Delta$, while the last one (i.e., the second breather $B_2$) has a larger gap $\sqrt{3} \Delta$. The mass $\Delta_n$ of the $n$-th breather is determined as 
	\begin{equation}
		\Delta_n = 2 \Delta \sin\left(\frac{\pi \xi n}{2}\right) = 2 \Delta \sin(\pi n/6)
	\end{equation}
	so that $\Delta_1 = \Delta$ (mass of the first breather coincides with that of the soliton and antisoliton), $\Delta_2 = \sqrt{3}\Delta$ (mass of the second breather) and 
	$\Delta_3 = 2\Delta$ (mass of the third breather, coincides with the boundary of the soliton-antisoliton continuum).
	
	And the value of the gap $\Delta$ is
	\ifdefined\details  \textcolor{red}{[Eq. (3.55) E-K + external field] } \fi
	\begin{align}
		\frac{\Delta}{J}= \sqrt{\pi}
		\left(\frac{v}{\pi J /2}\right)
		\frac{\Gamma(\frac{1}{6})}{\Gamma(\frac{2}{3})}\left[c \left(\frac{\pi J /2}{v}\right) \frac{\Gamma(\frac{3}{4})}{\Gamma(\frac{1}{4})}\frac{h+h_{\rm eff}}{J}\right]^{2/3}
	\end{align}
	Because $v= \pi J/2$ and $c=1/2$, we have
	\begin{align}
		\frac{\Delta}{J}= \frac{3^{3/4}\pi^{1/2}}{2}
		\gamma_1^{1/2}\left(\frac{h+h_{\rm eff}}{J}\right)^{2/3}
	\end{align}
	Here, we substitute $h_{\rm eff}$ with the solution of Eq.~\eqref{eq:heff:LZ/:part1}
	\begin{align}
		\frac{\Delta}{J}= \frac{3^{3/4}\pi^{1/2}}{2}
		\gamma_1^{1/2}
		\left(\frac{h}{J}\right)^{2/3}\left[1+\frac{3}{x}\cos\left(\frac{\arccos x}{3}\right)\right]^{2/3}
		\label{eq:gap:general:LZ}
	\end{align}
	where the value of $x$ can be computed via Eq.~\eqref{eq:heff:LZ/:part2}. This formula gives the gap of excitations, as a function of control parameters $h$ and $g$. Most importantly, here we define two quantities $\Delta_0$ and $b_h$.
	$\Delta_0$ is the gap value in the absence of external field
	\begin{align}
		\frac{\Delta_0}{J}=\frac{\Delta(h=0)}{J}=\frac{3^{3/4}\pi^{1/2}}{2} \gamma_1^{3/2} g
		\label{eq:Delta:0:LZ}
	\end{align}
	where $\gamma_1 \approx 1.215340$ as defined in Eq.~\eqref{eq:SI:gamma:1}.
	$b_h$ is a dimensionless quantity characterizing the field dependence
	$b_h=\left(\frac{\Delta}{\Delta_0} \right)^2-1$, where, $\Delta$ is the the value of the gap in the presence of an external field ($h$) and it is normalized by the zero-field gap value ($\Delta_0$). By definition, $b_h$ is a function of the field $h$ and its value vanishes at $h=0$. From Eq.~\eqref{eq:gap:general:LZ}, it is easy to check that
	\begin{align}
		b_h=\frac{(h/J)^{4/3}}{\gamma_1^2 g^2}\left[1+\frac{3}{x}\cos\left(\frac{\arccos x}{3}\right)\right]^{4/3}-1
		\label{eq:b0:LZ}
	\end{align}
	with $x$ defined in Eq.~\eqref{eq:heff:LZ/:part2}
	
	The spin-spin and bond-bond correlation functions are
	\begin{align}
		\left<S^x_i(t)S^x_j(t')\right>&=\left<S^y_i(t)S^y_j(t')\right>\simeq c^2 (-1)^{i-j} \left<\sin\left (\frac{\Theta(x_i, t)}{2}\right)\sin\left (\frac{\Theta(x_j, t')}{2}\right)\right>
		\\
		\left<S^z_i(t)S^z_j(t')\right>&\simeq c^2 (-1)^{i-j} \left<\cos\left (\frac{\Phi(x_i, t)}{2}\right)\cos\left (\frac{\Phi(x_j, t')}{2}\right)\right>
		\\
		\left<B_i(t)B_j(t')\right>& \propto \left<\sin\left (\frac{\Phi(x_i, t)}{2}\right)\sin\left (\frac{\Phi(x_j, t')}{2}\right)\right>
	\end{align}
	where $x_i=i a_0$ and $x_j=j a_0$ are the spatial coordinate of the the spin/bond. It is worthwhile to emphasize here that in the AFM phase ($m_0\ne 0$) with an order parameter along the $S^z$ direction, although the bond operator $B$ and $S^z$ share the same symmetry, the bond operator spectrum is characterized the same mass as that of  $S^x$ and $S^y$ operators, and is different from that of $S^z$. This fact will lead to a key consequence. As will be shown below, it implies that the $B-B$ correlator and the $S^z$-$S^z$ correlator are probing different excitations in quasi-1D systems. This is in sharp contrast to higher dimensions, where the two correlators can both be used to probe the amplitude mode due to their identical symmetry.
	
	These 1D correlation functions can be calculated. For $k_x$ near $\pi$, we have 
	\begin{align}
		\tilde{\chi}_{S^x}(\omega,k_x)&=\frac{Z_1}{\omega^2-v^2 (k_x-\pi)^2 - \Delta^2+ i \delta} + \textrm{incoherent background}
		\\
		\tilde{\chi}_{S^z}(\omega,k_x)&=\frac{Z_2}{\omega^2-v^2 (k_x-\pi)^2 - 3 \Delta^2 + i \delta}+ \textrm{incoherent background}
		\\
		\tilde{\chi}_{B}(\omega,k_x)&=\frac{Z_3}{\omega^2-v^2 (k_x-\pi)^2 - \Delta^2+ i \delta}+ \textrm{incoherent background}
	\end{align}
	where $\Delta$ is the soliton energy [Eq.~\eqref{eq:gap:general:LZ}]. If the external field is turned off ($h=0$), $\Delta$ recovers the zero-field value $\Delta_0$, as shown in Eq.~\eqref{eq:Delta:0:LZ}. Here, we only shows the coherent modes, while the incoherent background (the continuum) are ignored. Because the continuum only appear for $\omega\ge 2\Delta$ while the frequency of the coherent modes ($\omega=\Delta$ and $\omega=\sqrt{3}\Delta$) are all below $2\Delta$, the continuum can be ignored as far as these low-energy excitations are concerned.  As already mentioned above, the coherent modes of $S^x$ and the bond operator shares the same dispersion and energy gap ($\Delta$), while $S^z$ mode have a larger gap $\sqrt{3} \Delta$.
	
	\subsection{Random Phase Approximation and 2D Susceptibilities}
	In the previous section, we presented the dynamical susceptibilities of the spin/bond operators ($\tilde \chi$). In this section, 2D dynamical susceptibility will be computed via a random phase approximation (RPA). To distinguish 1D and 2D dynamical susceptibilities, 1D and 2D susceptibilities will be represented by $\tilde \chi$ and $\chi$ (with and without tilde on top) respectively.
	
	Within the RPA approximation, 2D dynamical susceptibilities for $S^x$ and $S^z$ are
	\begin{align}
		\chi_{S^x}&=\frac{1}{(\tilde{\chi}_{S^x})^{-1}-2 J'(k_y)}
		\\
		\chi_{S^z}&=\frac{1}{(\tilde{\chi}_{S^z})^{-1}-2 J'(k_y)}
	\end{align}
	where $J'(k_y)=J_\perp \cos k_y$ is the Fourier transform of the inter-chain coupling. As for the bond-bond correlation function, because the interchain bond-bond coupling comes from second-order (and higher-order) perturbation in $g\ll 1$, the interchain coupling constants scales as $J_\perp^2/J$ and thus is small for weak $J_\perp$, $J_\perp^2/J \ll J_\perp$ \cite{hikihara2010}. As a result, to the leading order, the bond-bond correlation does not receive any corrections within the RPA approximation, and therefore
	\begin{align}
		\chi_{B}&=\tilde{\chi}_{B}
	\end{align}
	
	In summary, we find
	\begin{align}
		\label{eq:chi-Sx-rpa}
		\chi_{S^x}(\omega,\mathbf{k})&=\chi_{S^y}(\omega,\mathbf{k})=\frac{Z_1}{\omega^2-v^2 (k_x-\pi)^2 - (\Delta^2+2 Z_1 J_\perp \cos k_y)+ i \delta}
		\\
		\label{eq:chi-Sz-rpa}
		\chi_{S^z}(\omega,\mathbf{k})&=\frac{Z_2}{\omega^2-v^2 (k_x-\pi)^2 - (3 \Delta^2+2 Z_2 J_\perp \cos k_y)+ i \delta}
		\\
		\label{eq:chi-B-rpa}
		\chi_{B}(\omega,\mathbf{k})&=\frac{Z_3}{\omega^2-v^2 (k_x-\pi)^2 - \Delta^2+ i \delta}
	\end{align}
	
	From the Goldstone theorem, we know that at $h=0$, in the SSB phase the phase fluctuations mode $\chi_{S^x}$ shall be gapless at $k_y=\pi$. This condition fixes the value of $Z_1$
	\begin{align}
		\Delta_0=\sqrt{2 Z_1 J_{\perp}}.
	\end{align}
	Because 1D exact methods fix the ratio between $Z_1$ and $Z_2$, $Z_2/Z_1\approx 0.491309$~\cite{Quasi-one-dimensional1997Essler}, the value of $Z_2$ is also determined. As for $Z_3$, because its value doesn't change the mode frequency, it will not be evaluated here.
	
	As a result, the dispersion of these three modes are
	\begin{align}
		\omega_{S^x}&=\omega_{S^y}= \Delta_0 \sqrt{1+ b_h + \cos k_y + \frac{v^2 (k_x-\pi)^2}{\Delta_0^2}}
		\\
		\omega_{S^z}&= \Delta_0 \sqrt{3(1+ b_h) + \frac{Z_2}{Z_1}\cos k_y+ \frac{v^2 (k_x-\pi)^2}{\Delta_0^2}}
		\\
		\omega_{B}&= \Delta_0\sqrt{1+b_h + \frac{v^2 (k_x-\pi)^2}{\Delta_0^2}}
	\end{align}
	where $\Delta_0$ is the value of $\Delta$ in the absence of external field [Eq.~\eqref{eq:Delta:0:LZ}] and the dimensionless parameter $b_h$ is function of the field strength, defined in Eq.~\eqref{eq:b0:LZ}.
	The ratio $Z_2/Z_1\approx0.491309$. The velocity along the chain is $v=\pi J/2$ as shown above.

	At $k_x=\pi$ and in the absence of external field ($h=0$ and thus $b_h=0$), the dispersion relations become
	\begin{align}
		\omega_{S^x}&=\omega_{S^y}= \Delta_0 \sqrt{1+ \cos k_y}
		\\
		\omega_{S^z}&= \Delta_0 \sqrt{3+ \frac{Z_2}{Z_1}\cos k_y}
		\\
		\omega_{S^B}&= \Delta_0
	\end{align}
	And it is easy to check that near $k_x=\pi$, for any values of $k_y$, the frequency of these modes never reach $2\Delta$, which is the onset frequency, above which the 1D incoherent continuum start to arise. This fact justifies the procedure above, where we dropped the incoherent background in $\tilde{\chi}$. In addition, this observation also implies that these modes are lightly damped at small $J_{\perp}$ (i.e., the damping only comes from interchain fluctuations), which is the key reason why they leads to clear peaks in the spectrum functions.

	\section{Amplitude and phase modes from QMC measurements}
	
	\begin{figure*}[htp]
		\centering
		\includegraphics[width=\textwidth]{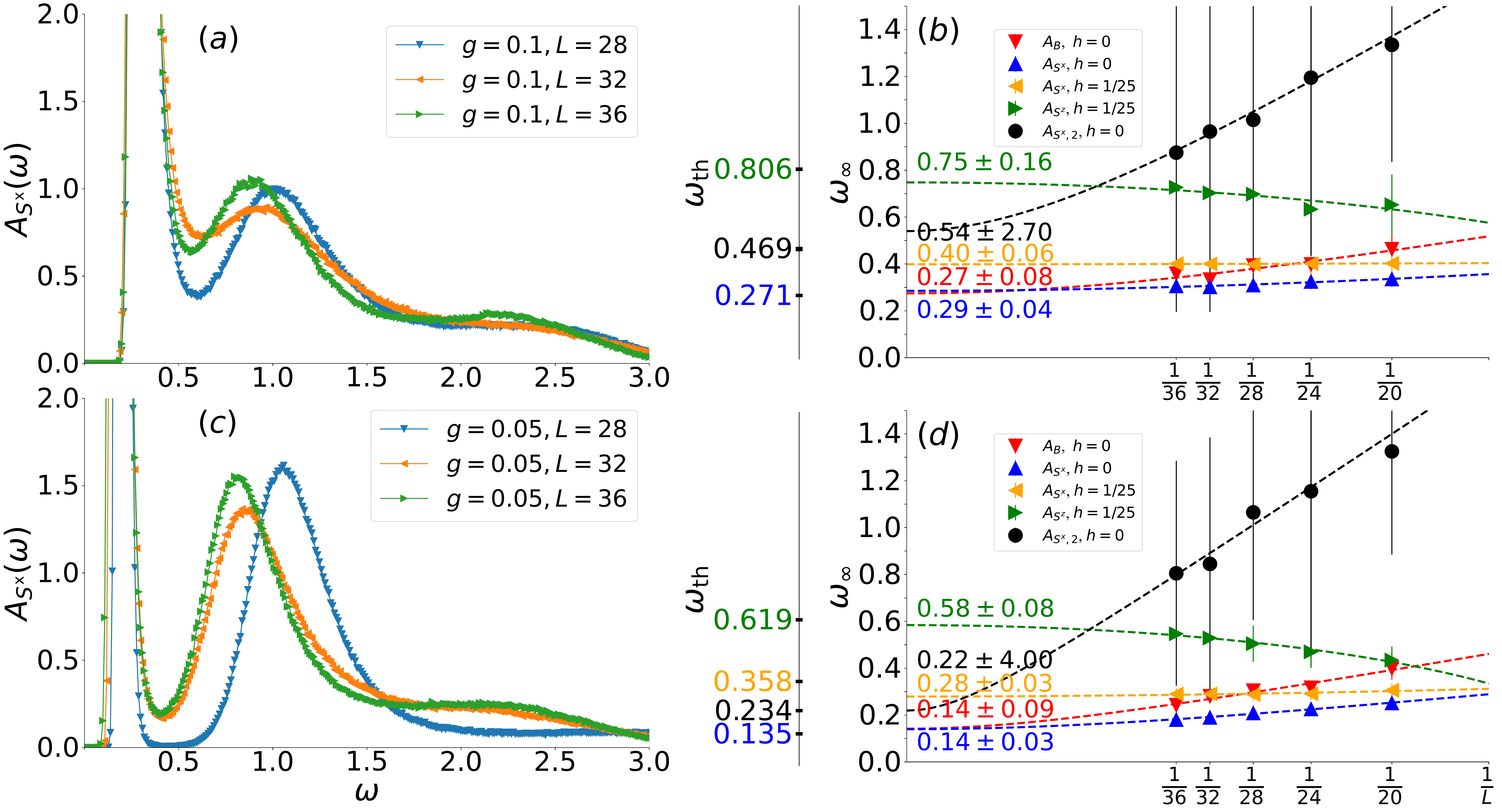}
		\caption{Frequency dependence of the spectral functions at $\mathbf{k}=(\pi,\pi/2)$ for $A_{S^x}(\omega)$ with $h=0$ for (a) at $g=0.1$ and (c) at $g=0.05$. And finite-size analysis for (b) $g=0.1$ and (d) $g=0.05$ at the same momentum point. The vertical $\omega_\infty$-axis shows extrapolation of the numerical data to the $L=\infty$ limit. Values on the vertical $\omega$-axis mark analytical predictions for the peak frequencies of different modes. See text for details.}	
		\label{fig:figS4}
	\end{figure*}

	In this section, we show that in the absence of the staggered pinning field ($h$), the spin-spin correlation measured in QMC simulations shall reveal information about both phase and amplitude fluctuations.
	
	Because spontaneous symmetry breaking only arises in the thermodynamic limit, while QMC simulations are performed with finite-size systems, all measurements in such finite-size simulations shall exhibit the full symmetry of the Hamiltonian (without spontaneous symmetry breaking) in both ordered and disordered phase. For example, in a model with Heisenberg spins, (i.e. the model studied in this manuscript), the spin-spin correlation function measured in QMC simulations will always exhibit the SO(3) symmetry, unless a pinning field $h$ is introduced to explicitly break this symmetry.
	
	This absence of spontaneous symmetry breaking is a finite-size effect. Same as most other finite-size effects, it implies that finite-size analysis and extrapolation are required, in order to access the thermodynamic limit. However, as will be shown below, this particular finite-size effect, i.e., the absence of spontaneous symmetry breaking, provides a tool to access both phase and amplitude fluctuations in one single correlation function, in contrast to the thermodynamic limit, where information about the amplitude and phase modes are encoded in different components of spin fluctuations, parallel and perpendicular to the direction of the magnetic order respectively.
	
	In the ordered phase, because of the finite size, QMC simulations will sample the entire phase space, instead of just one ordered state. When the system size is large enough (but still finite), this effect can be characterized via the density matrix, which takes the following form to the leading order
	\begin{align}
		\rho = \sum_{\vec{n}} \rho_{\vec{n}} + \ldots,
		\label{SM:eq:density:matrix}
	\end{align}
	Here, $\rho$ is the density matrix of QMC simulations, which samples all possible configurations. $\rho_{\vec{n}}$ is the restricted density matrix, limited to sample only configurations around one of the ordered states, whose order parameter is aligned to the $\vec{n}$ direction, with $\vec{n}$ being a 3D unit vector. The sum here is performed over all possible direction of $\vec{n}$. The $\ldots$ represent higher order terms, e.g. cross terms between two different ordered states, which vanishes in the thermodynamic limit. For simplicity, here we will ignore all such higher order terms, and only focus on the leading contribution $\sum_{\vec{n}} \rho_{\vec{n}}$. 
	
	For a real physical system in the thermodynamic limit, only a single $\rho_{\vec{n}}$ will be selected in the ordered phase, due to spontaneous symmetry breaking. However, for finite-size systems, all $\rho_{\vec{n}}$'s will contribute to the ensemble average with the same probability, as shown in Eq.~\eqref{SM:eq:density:matrix}. As a result, if we measure a correlation function in the ordered phase, e.g., $\langle S^x_i S^x_j \rangle$, it shall get signals from all possible symmetry breaking states
	\begin{align}
		\langle S^x_i S^x_j \rangle_{\textrm{QMC}}=\frac{1}{Z}\sum_{\vec{n}} \textrm{tr} \left(\rho_{\vec{n}} S^x_i S^x_j\right),
		\label{SM:eq:ensemble:average:1}
	\end{align}
	where the normalization factor $Z$ is the partition function.  Due to the SO(3) rotational symmetry, we can rewrite this formula as
	\begin{align}
		\langle S^x_i S^x_j \rangle_{\textrm{QMC}}=\frac{1}{Z}\sum_{\vec{n}} \textrm{tr} \left(\rho_{\vec{z}} S^{\vec{n}}_i S^{\vec{n}}_j\right),
		\label{SM:eq:ensemble:average:2}
	\end{align}
	In Eq.~\eqref{SM:eq:ensemble:average:1}, we fix the direction of the spin ($S^x$) and let the direction of the order parameter to rotate and to explore all possible solid angle. In Eq.~\eqref{SM:eq:ensemble:average:2}, instead, we rotate the direction of the spin $S^{\vec{n}}$ with a fixed direction for the order parameter ($\rho_{\vec{z}}$, i.e. the order parameter is along the z direction). Due to the rotational symmetry, these two options are fully equivalent, once we sum over all possible solid angle $\sum_{\vec{n}}$.
	
	With proper normalization factor, $\sum_{\vec{n}} S^{\vec{n}}_i S^{\vec{n}}_j =(S^x_i S^x_j + S^y_i S^y_j + S^z_i S^z_j)/3$, we therefore get
	\begin{align}
		\langle S^x_i S^x_j \rangle_{\textrm{QMC}}=\frac{1}{3} \left(\langle S^x_i S^x_j \rangle_{z} +\langle S^y_i S^y_j \rangle_{z} + \langle S^z_i S^z_j \rangle_{z} \right),
		\label{SM:eq:ensemble:average:3}
	\end{align}
	where $\langle \ldots \rangle_{\textrm{QMC}}$ represents the expectation value obtained from QMC simulations with contributions from all possible ordered states, and $\langle \ldots \rangle_{z}$ is the expectation value from only one ordered state with the order parameter aligned to the $z$ direction. Here, it is easy to realize that $\langle S^x_i S^x_j \rangle_{z}$  and $\langle S^y_i S^y_j \rangle_{z}$ measure phase fluctuations perpendicular to the order parameter, while $\langle S^z_i S^z_j \rangle_{z}$ gives amplitude fluctuations with spin along the order parameter direction. Thus, in QMC simulations, $\langle S^x_i S^x_j \rangle_{\textrm{QMC}}$ shall contain signals from both channels, phase and amplitude. This is the reason why we can see both these two modes in this one single correlation function in QMC. 
	
	For a regular 2D system with magnetic order, although $\langle S^x_i S^x_j \rangle_{\textrm{QMC}}$ in principle could detect both phase and amplitude fluctuations, the amplitude contribution will typically be buried in the incoherent background of phase fluctuations. This is because in a typical 2D magnet, amplitude fluctuations are strongly damped and the characteristic frequency range of such strongly damped fluctuations coincide with incoherent background of the phase fluctuations. As a result, we may add contributions from both phase and amplitude fluctuations in Eq.~\eqref{SM:eq:ensemble:average:3}, no clear signature of the amplitude part shall be observed. This is in good agreement with existing QMC studies as well as the 2D limit with $g=0.5$ showed in the main text.  However, in the quasi-1D limit, the sharp and strong amplitude mode shall give a clear feature in the QMC measurement,  as shown in the main text. This physics (where one observe both the phase and amplitude mode via $\langle \ldots \rangle_{\textrm{QMC}}$) is highly generic and in principle it shall arise in any systems where a strong and under-damped amplitude mode emerges.
	
	In practice, in order to detect the amplitude mode via $\langle S^x_i S^x_j \rangle_{\textrm{QMC}}$, it needs to first convert imaginary time (Matsubara frequency) QMC measurements to real time (real frequency) via analytical continuation. Then, in the real-frequency spectral function $A_{S^x}$, one shall observe two peaks: the phase mode at a lower frequency and the amplitude mode at a higher frequency. This is exactly what observed in our simulations at quasi 1D (small $g$) as shown in Fig.~\ref{fig:figS4}, where we plot $A_{S^x}$ at $k=(\pi,\pi/2)$ at $h=0$. From the figure, two clear features (peaks) emerge in the quasi-1D limit (small $g$).
	
	From the QMC data, the second, higher-energy peak in $A_{S^x}$ (denoted as $A_{S^x,2}$ in Fig.~\ref{fig:figS4}) is much broader than the lowest peak. This is mostly due to inherent difficulty of analytical continuation. By definition, a real-frequency mode at frequency $\omega$ manifests itself as an exponential decay in imaginary time with decay rate $\omega$, i.e., real frequency oscillations $A e^{-i \omega t}$ implying a decay in the corresponding imaginary-time correlation function $A e^{-\omega \tau}$. Thus, if we have multiple modes in real time/frequency, the imaginary-time correlation function shall exhibit multiple exponential decays, i.e. 
	\begin{align}
		A_1 e^{-i \omega_1 t}+ A_2 e^{-i \omega_2 t}+\ldots    \Leftrightarrow A_1 e^{-\omega_1 \tau}+ A_2 e^{- \omega_2  \tau}+\ldots
	\end{align}
	In the presence of multiple exponential decay rates, the mode with the lowest frequency (i.e. the slowest decay rate in imaginary time) shall give dominate contributions, while contributions from high-frequency modes are suppressed and sub-leading due to the larger decay rates. As a result, the 2nd (i.e. the higher frequency $A_{S^x,2}$) mode suffers more from noise and numerical uncertainty in both the simulations and the procedure of analytical continuation, as we convert imaginary time QMC data to real frequency. This is the main reason why the error-bar and uncertainty associated with the high-energy (amplitude) mode $A_{S^x,2}$ is much larger, in comparison with the low-energy one.  This problem and numerical error can be efficiently suppressed by apply a staggered pinning field ($h$). As shown in the main text, the pinning field pins the direction of the order parameter to $z$, and thus the phase and amplitude modes now are diverted into two different correlation functions $\langle S_x S_x \rangle$ and $\langle S_z S_z \rangle$, respectively. As a result, in $\langle S_z S_z \rangle$, there is no phase mode and the amplitude mode becomes the lowest frequency mode with dominant contribution to the imaginary-time QMC data, which leads to a much sharper amplitude-mode peak in comparison with $h=0$ case.
	
	It is also worthwhile to point out that in the theory picture above, we ignored higher order contributions from  finite-size effects in Eq.~\eqref{SM:eq:density:matrix}. Thus, finite-size analysis and extrapolation are still needed to directly compare theory predictions and numerical experiments in the thermodynamic limit. In Fig.~\ref{fig:figS4}, we plot $A_{S^x}$ at $k=(\pi,\pi/2)$ at $h=0$, where two features (peaks) emerge in the quasi-1D limit (small $g$). Upon increasing the system size $L$, the frequency of the 2nd peak decreases. Association of the 2nd peak with the 2nd breather of the sine-Gordon model implies that its energy should reach $\sqrt{3} \Delta_0$ in the $L=\infty$ limit. Our numerical data is consistent with this prediction, even though large errors (estimated as a half-widths of the 2nd peak) make a more definite conclusion impossible.
	
	It is worth reiterating again that our assignment of the 2nd peak to the 2nd breather is specific to $h=0$ spectral function $A_{S^x}$. It relies on the spin rotational symmetry of the $L\times L$ cluster at finite temperature. As a result, spin spectral function $A_{S^x}$ probes all spin components, much like an inelastic neutron scattering experiments with non-polarized neutrons which probe all components of the spin correlation function. Adding finite staggered $h$ changes the situation completely. Now the 2nd breather can only be observed in the $A_{S^z}$ spectral function, Fig.~\bl{2} (h) and (l). At the same time, solitons and anti-solitons are only present in $A_{S^x}$, Fig.~\bl{2} (g) and (k). Also, keep in mind that finite $h=1/25$ changes the soliton gap $\Delta_0$ to $\Delta_0 \sqrt{1+b_h}$, and correspondingly the energy of the 2nd breather becomes $\sqrt{3} \Delta_0 \sqrt{1+b_h}$.
	
\end{widetext}

\end{document}